\documentclass[aps,prb,twocolumn,superscriptaddress,10pt]{revtex4-1}

\usepackage{graphicx}				
\usepackage{epsfig}
\usepackage{amsmath,amsthm} 
\usepackage{color}
\usepackage{verbatim}				
\usepackage{ulem}
\begin{document}

\title{Valence-transition-induced changes of the electronic structure in EuPd$_2$Si$_2$}

\author{O. Fedchenko}
\affiliation{Institut f\"{u}r Physik, Johannes Gutenberg-Universit\"{a}t, Staudingerweg 7, D-55128 Mainz, Germany}

\author{Y.-J. Song}
\affiliation{Institut f{\"u}r Theoretische Physik, Goethe-Universit{\"a}t Frankfurt, Max-von-Laue Strasse 1, 60438 Frankfurt am Main, Germany}

\author{O. Tkach}
\affiliation{Institut f\"{u}r Physik, Johannes Gutenberg-Universit\"{a}t, Staudingerweg 7, D-55128 Mainz, Germany}

\author{Y. Lytvynenko}
\affiliation{Institut f\"{u}r Physik, Johannes Gutenberg-Universit\"{a}t, Staudingerweg 7, D-55128 Mainz, Germany}
\affiliation{Institute of Magnetism of the NAS and MES of Ukraine, 03142 Kyiv, Ukraine}

\author{S. V. Chernov}
\affiliation{Deutsches Elektronen-Synchrotron DESY, 22607 Hamburg, Germany}

\author{A. Gloskovskii}
\affiliation{Deutsches Elektronen-Synchrotron DESY, 22607 Hamburg, Germany}
\author{C. Schlueter}
\affiliation{Deutsches Elektronen-Synchrotron DESY, 22607 Hamburg, Germany}

\author{M. Peters}
\affiliation{Physikalisches Institut, Goethe Universit\"{a}t Frankfurt, Max-von-Laue-Strasse 1, 60438 Frankfurt am Main, Germany}

\author{K. Kliemt}
\affiliation{Physikalisches Institut, Goethe Universit\"{a}t Frankfurt, Max-von-Laue-Strasse 1, 60438 Frankfurt am Main, Germany}

\author{C. Krellner}
\affiliation{Physikalisches Institut, Goethe Universit\"{a}t Frankfurt, Max-von-Laue-Strasse 1, 60438 Frankfurt am Main, Germany}

\author{R. Valent\'\i}
\affiliation{Institut f{\"u}r Theoretische Physik, Goethe-Universit{\"a}t Frankfurt, Max-von-Laue Strasse 1, 60438 Frankfurt am Main, Germany}

\author{G. Sch{\"o}nhense}
\affiliation{Institut f\"{u}r Physik, Johannes Gutenberg-Universit\"{a}t, Staudingerweg 7, D-55128 Mainz, Germany}

\author{H.J. Elmers}\email{elmers@uni-mainz.de}
\affiliation{Institut f\"{u}r Physik, Johannes Gutenberg-Universit\"{a}t, Staudingerweg 7, D-55128 Mainz, Germany}


\date{\today}

\begin{abstract}

We present results of hard X-ray angle-resolved photoemission spectroscopy and photoemission diffraction measurements performed on high-quality single crystals of the valence transition compound EuPd$_2$Si$_2$ for temperatures 25~K $\leq$ T $\leq$ 300~K. 
At low temperatures we observe a Eu $4f$ valence $v=2.5$, 
which decreases to $v=2.1$ for temperatures above the valence transition around $T_V \approx 160$~K. The experimental valence numbers resulting from an evaluation of the Eu(III)/Eu(II) $3d$ core levels, are used for calculating band structures using density functional theory.
The valence transition significantly changes the band structure as determined by angle-resolved photoemission spectroscopy. In particular, the Eu $5d$ valence bands are shifted to lower binding energies with increasing Eu $4f$ occupancy. To a lesser extent,  bands derived from the Si $3p$ and Pd $4d$ orbitals are also affected. This observation suggests a partial charge transfer between Eu and Pd/Si sites. Comparison with {\it ab-initio} theory shows a good agreement with experiment, in particular concerning the unequal band shift with increasing Eu $4f$ occupancy.

\end{abstract}

\maketitle

\section{Introduction}

A large variety of interesting
phenomena, such as the Kondo effect and the appearance of
heavy fermion features, quantum criticality, unconventional
superconductivity, exotic magnetism, and non-trivial topological
phases have been observed
in metallic compounds containing rare earth elements~\cite{Wirth2016,Rau2019,Pfleiderer2009,Stewart2001,Onuki2020,Onuki2016,Xu2020,Rahn2018,Schulz2019}. 

Eu is particularly known to exhibit valence transitions due to the small energy difference between the divalent and trivalent states of the Eu ion,
as has been observed
in the tetragonal compounds
EuPd$_2$Si$_2$~\cite{Sampathkumaran1981}, EuCu$_2$Si$_2$~\cite{Bauminger1973,Patil1993}, and EuIr$_2$Si$_2$~\cite{Chevalier1986,Seiro2019,Schulz2019}. 
These valence transitions also occur under the
application of pressure in the tetragonal antiferromagnets
EuRh$_2$Si$_2$~\cite{Mitsuda2012}, EuNi$_2$Ge$_2$~\cite{Nakamura2012}, and EuCo$_2$Ge$_2$~\cite{Dionicio2006}. 
Alternatively, chemical pressure by partial replacement of constituents has been used
to tune the valence transition~\cite{Ichiki2017}.
Recently, the valence dynamics has been investigated using time-resolved absorption spectroscopy~\cite{Yokoyama2019}.

\begin{figure}
\includegraphics[width=\columnwidth]{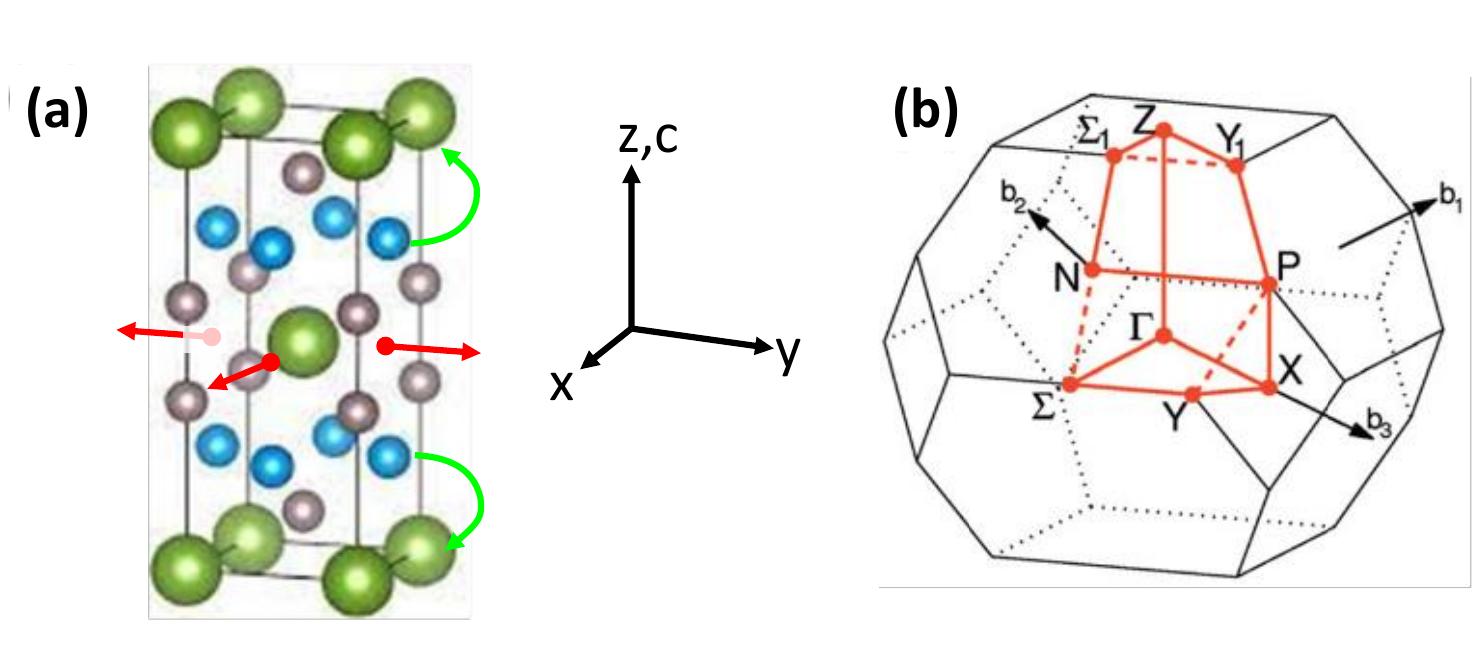}
\caption{\label{Fig1} 
(a) Sketch of the direct space crystal structure of
EuPd$_2$Si$_2$ and coordinate system used in this article.
Red arrows denote the lattice expansion with increasing temperature in the ab-plane caused by the valence transition.
Green arrows denote the partial charge transfer from the
Si/Pd-planes to the Eu crystal plane.
(b) Corresponding Brillouin zone in reciprocal space indicating the high-symmetry points.
}
\end{figure}

EuPd$_2$Si$_2$ has attracted particular scientific attention because the Eu ions in this material exhibit a substantial valence transition from a 
non-integer value close to a magnetic Eu(II) state to a non-integer value close to a nonmagnetic Eu(III) state upon decreasing
the temperature~\cite{Onuki2016}. 
The valence transition is associated with a significant change in the magnetic properties. Eu(II) has a large magnetic moment associated with localized $4f$ -electrons, whereas Eu(III) has a zero moment ground state~\cite{Sampathkumaran1981}.
In this compound, the valence transition near 150~K is also linked to a substantial volume reduction from the Eu(II) state at room temperature to the dominant Eu(III) state below the valence transition ambient pressure. 
The valence transition can be tuned by temperature~\cite{Sampathkumaran1981}, pressure~\cite{Adams1991}, or high magnetic fields~\cite{Mitsuda1997}.
Also, small particle sizes influence the valence transition~\cite{Iyer2018}.
When going through the valence transition, the lattice parameter $a$ is strongly reduced by about 2\%~\cite{Onuki2016}, while the lattice parameter $c$ remains almost constant [Fig.~\ref{Fig1}(a)]. 
Due to this interplay between lattice and electronic degrees of freedom, the strain in epitaxial thin films has been shown to suppress the valence transition to temperatures below 20~K~\cite{Koelsch2022}. Below this temperature, electronic transport measurements indicate antiferromagnetic ordering.

While the changes in the lattice structure during the valence transition are well known, the corresponding changes in the electronic structure as observed by surface sensitive photoemission spectroscopy were found to be obscured by the different behavior of the material at the surface~\cite{Maartensson1982,Mimura2004,Mimura2004a}.

The lack of nearest neighbor atoms at the surface (100) surface allows expansion to counteract the valence transition at low temperatures. The high temperature divalent state remains partially at the surface even at lower temperatures, as has been found by photoemission spectroscopy in the vacuum ultraviolet and soft X-ray regions~\cite{Mimura2004,Mimura2004a}. 
In contrast, hard X-ray photoemission spectroscopy reveals the bulk electronic structure~\cite{Gray2011}. 

The Anderson model analysis ascribes the drastic
valence change to the hybridization of the Eu $4f$ electrons
with the conduction electrons. Density Functional Theory calculations~\cite{Song2023} support this description. To experimentally validate these studies we performed
hard X-ray angle-resolved photoemission spectroscopy experiments.
We observe changes in the spectral shape of the trivalent Eu(III) 3$d$ spectra at the valence transition as well as significant variations of the valence band structure in agreement with density functional calculations.

\section{Experimental}

Single crystals of the compound EuPd$_2$Si$_2$
were synthesized
using the Czochralski method from an Eu-rich levitating melt as described in Ref~\onlinecite{Kliemt2022}.
The single crystals were bonded to the sample holder using epoxy resin 
with the (001) plane
parallel to the sample holder plate. 
Before being transferred into the vacuum chamber, the crystals were cleaved using
a wire cutter. 
The samples are inserted into a He-cooled (25~K) sample stage on a high-precision 6-axis hexapod manipulator of the time-of-flight momentum microscope. 

The experiments were performed at beamline P22 of the storage ring PETRA III at 
DESY in Hamburg (Germany)~\cite{Schlueter2019}. 
The photon energy was set to 3.4~keV in a spot of about 
$10\times 10$~$\mu$m$^2$ using a Si(220) double-crystal monochromator. 
At 3.4~keV the total energy resolution is determined by the photon band width of 100~meV. 

A major advantage of angle-resolved photoelectron spectroscopy in the hard X-ray range is the significant increase in the inelastic mean free path of the escaping photoelectrons. 
Therefore, the present results represent true bulk properties.
This is essential, because the valency of Eu at the surface of valence fluctuating systems is divalent and the valence crossover cannot be studied~\cite{Mimura2004,Mimura2004a,Schulz2019}. 

To overcome the challenges of low cross-section and low signal-to-background
ratio in the hard X-ray regime, we applied 
time-of-flight momentum microscopy~\cite{Medjanik2017}, 
which allows highly-efficiency three-dimensional data acquisition of the
photoelectron intensity $I(E_B,k_x,k_y)$ as a function of
binding energy, $E_B=-(E-E_F)$, and momentum. 
The large momentum field of view of 12~\AA$^{-1}$ simultaneously yields 
five adjacent Brillouin zones, where larger parallel momentum leads to
a decreasing perpendicular momentum. 
This allows us to resolve the perpendicular momentum $k_z$, covering
in one experimental run a range of $\Delta k_z=0.5G_{001}$,
as shown in Ref.~\cite{Agustsson2021}.
Noise reduction data processing limits the momentum resolution to $0.08$~\AA$^{-1}$ for the results presented below.

Data were acquired at 25~K and 300~K for 8 hours in each case.
Details of the data evaluation procedure are described in Refs.~\cite{Babenkov2019,Medjanik2019}.

\section{Experimental Results}

\begin{figure}
\includegraphics[width=0.95\columnwidth]{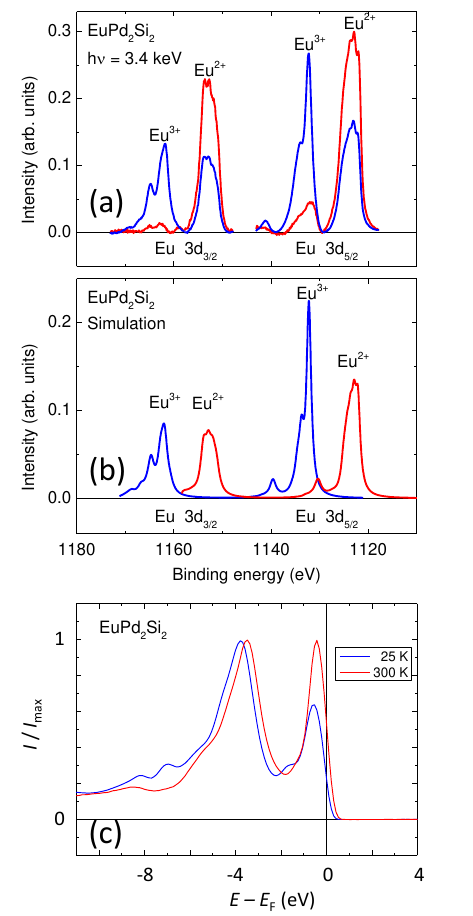}
\caption{\label{Fig2} 
(a) Temperature dependence of the Eu $3d$ core spectra obtained with a photon energy of 3.4~keV at 300~K (red) and 25~K
(blue) for a EuPd$_2$Si$_2$ single crystal. 
The mean valence number $n$ results is derived from the ratio of the areas of the corresponding
$3d$ core orbitals. The specified binding energies are referred to the Fermi level.
(b) Corresponding simulation using charge transfer multiplet calculations.
(c) Valence band energy distribution curves for 25~K and 300~K.
The photoemission intensity was integrated over a planar section of the full Brillouin zone.
}
\end{figure}

To quantify the valence number $v$ of the Eu ions above and below the valence transition we measured the Eu X-ray photoelectron spectra (XPS) at a photon energy of 3.4~keV as shown in Fig.~\ref{Fig2}(a). 
At this photon energy the inelastic mean free path of the photoelectrons is $\approx$ 4~nm~\cite{Seah1979} and one obtains bulk-related information rather than surface properties.
Due to the spin-orbit interaction
the Eu $3d$ spectrum is split into an Eu $3d_{5/2}$ and an Eu$3d_{3/2}$ component with a splitting energy of 30~eV.
Each component is further split by a chemical shift depending on the valence state, which is either  Eu(II) or Eu(III).
The ionization state splitting is 10~eV.
This large splitting allows a precise
determination of the mean valence from the corresponding
peak areas~\cite{Mimura2011}. 

Comparison of the high (300~K)
and low (25~K) temperature spectra shows a significant
change in the ratio between the Eu(II) and Eu(III) components.
An evaluation of the peak area ratio results in values for the 
valence number $v=2.1$ (occupation number $n=6.9$) at high temperature, corresponding to an
almost complete Eu(II) ionization state, and an increase to $v=2.5$ (occupation number $n=6.5$)
at low temperature, indicating a mixed Eu(II)/Eu(III) valence state.

Multiplet calculations~\cite{Haverkort2016} of the XPS spectra, taking into account interatomic correlation effects, show a very good agreement with the experimental spectra [Fig.~\ref{Fig2}(b)].
The Eu(III) spectra are split by 2~eV due to an atomic multiplet effect. For the Eu(II) spectra we observe an additional splitting that is caused by the inter-atomic exchange interaction. The Eu(II) ion carries a local magnetic moment of 6~$\mu_B$ localized in the $4f$ orbitals, while the magnetic moment of Eu(III) is zero.
The Eu $3d$ peaks are then split according to the magnetic quantum number of the total angular momentum. Thus, the different magnetic properties of Eu(II) and Eu(III) ions are directly visible in the XPS spectra. 

The valence band photoemission intensity measurements [shown in Fig.~\ref{Fig2}(c)]
show the density of states at low and high temperatures.
The spectra have been normalized to the maximum at 4~eV binding energy.
The sharp peak near the Fermi level is due to the
partially filled Eu(II) $4f$ states. The increase of this peak
reflects the increase of the Eu(II)/Eu(III) ratio at high temperature.
The Eu(III) $4f$ states cause the small peaks at 2~eV and 8~eV binding energy at low temperature, which disappear at high temperature.
Thus, also the valence spectra confirm the 
presence of an almost pure Eu(II) state at 300~K and a strong redistribution of the peak heights on cooling
below the valence transition temperature,
in agreement with measurements on polycrystalline EuPd$_2$Si$_2$ samples~\cite{Mimura2004}. 

\begin{figure}
\includegraphics[width=\columnwidth]{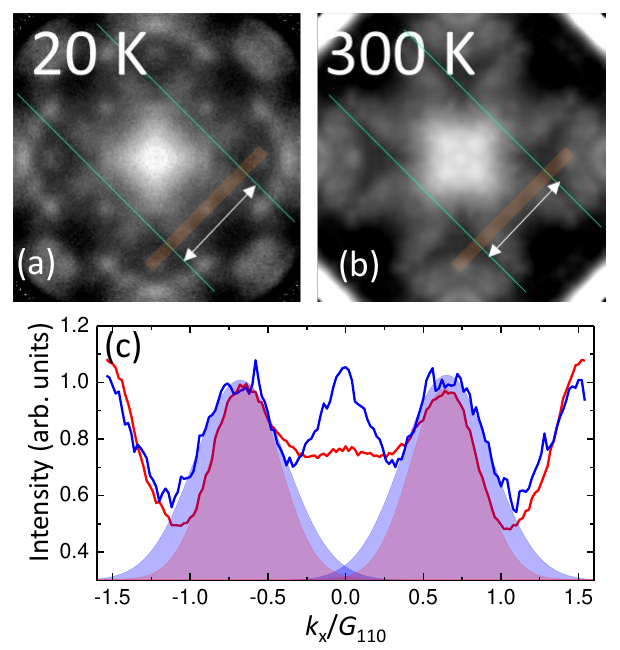}
\caption{\label{Fig3} 
(a) X-ray photoelectron diffraction patterns measured for a photon energy of 3.4~keV at the Eu(II) $3d_{5/2}$ core level at low temperature. Dark straight lines indicated by blue lines mark Kikuchi lines. The distance between a pair of Kikuchi lines corresponds to $2G_{110}$.
(b) Similar data measured at room temperature.
(c) Line profiles averaged along the yellow thick lines in (a) and (b) (Blue full line for low temperature and red full line for room temperature). Fits to Gaussian functions of two corresponding intensity maxima reveal an increase of the reciprocal lattice in the ab-plane by $(3\pm 1)$\% at low temperature as compared to room temperature.
}
\end{figure}

\begin{figure}
\includegraphics[width=\columnwidth]{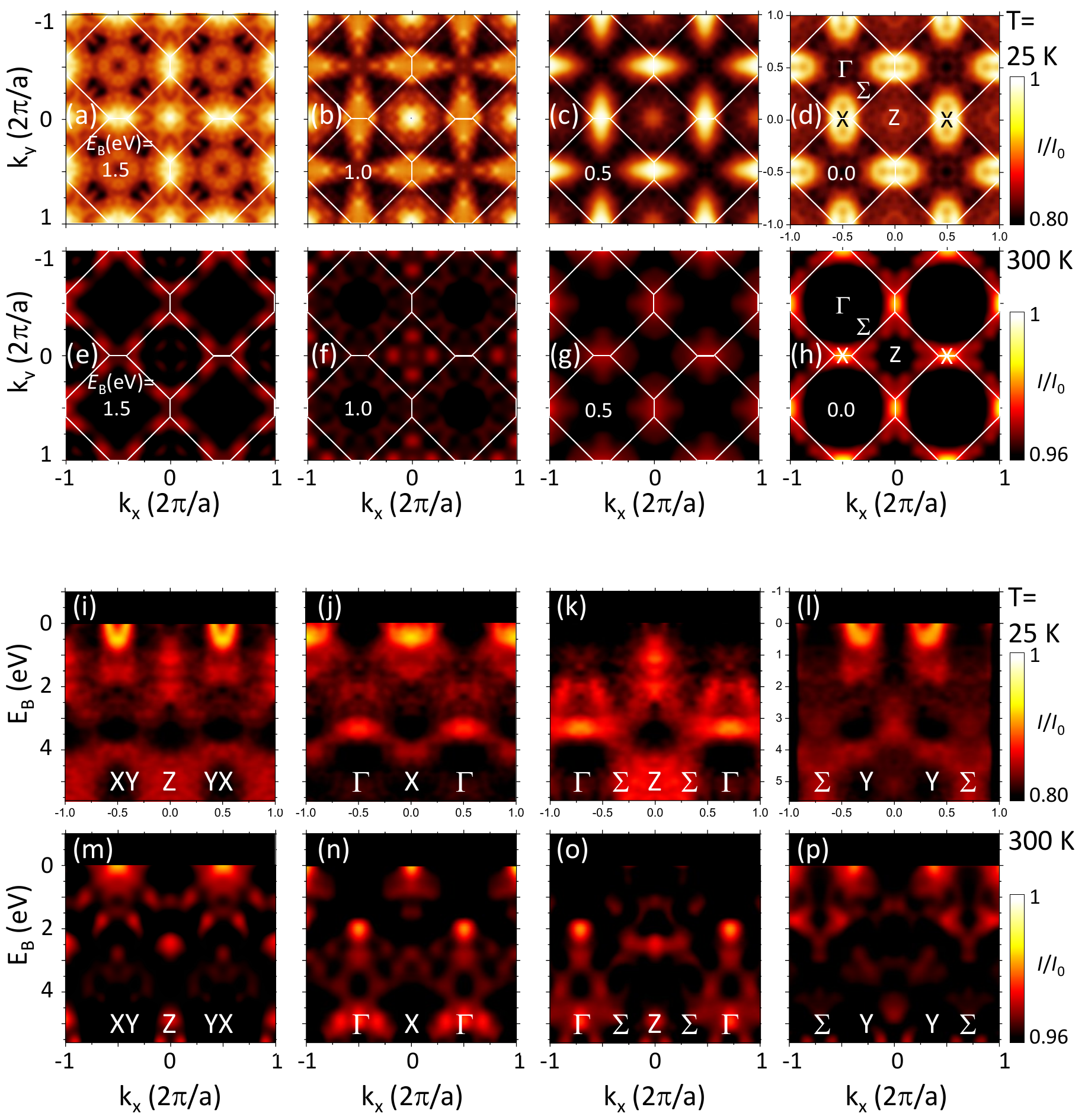}
\caption{\label{Fig4} 
(a-h) Constant energy maps of the photoemission intensity
$I(E_B,k_x,k_y)$ for the indicated binding energies $E_B$ measured at 25~K (a-d) and 300~K (e-h). The photon energy is 3.4~keV. The photoemission intensity has been symmetrized according to the crystal symmetry.
(i-p) Binding energy versus parallel momentum sections of the photoemission data array along the indicated high symmetry directions measured at 25~K (i-l) and 300~K (m-p).
The photoemission intensity is normalized on the 
valence band energy distribution curve shown in Fig.~\ref{Fig2}(c) and color coded on the indicated linear orange-hot scale.
}
\end{figure}


X-ray photoelectron diffraction (XPD) patterns measured for a photon energy of 3.4~keV at the Eu(II) $3d_{5/2}$ core level reveal changes in the lattice structure upon cooling beyond the valence transition temperature [see Fig.~\ref{Fig3}(a,b)]. Kikuchi lines indicate the in-plane reciprocal lattice vectors. A pair of parallel Kikuchi lines along the in-plane [110] directions is represented by straight lines in 
Fig.~\ref{Fig3}(a,b). The distance between these two Kikuchi lines is twice the reciprocal lattice vector $G_{110}$.
To quantify the in-plane lattice expansion, we compare the line profiles
over the Kikuchi lines as shown in Fig.~\ref{Fig3}(c).
A Gaussian function fit to the two prominent intensity maxima
leads to a reduction in $G_{110}$ of $3\pm 1$\% due to heating above the valence transition temperature. 
The reduction in the reciprocal space corresponds to an increase in the lattice in the direct space.
The observed change is consistent with previously observed changes of the lattice parameter using X-ray diffraction~\cite{Kliemt2022}.
In contrast to X-ray diffraction, XPD probes the same sample volume as the angle-resolved photoemission spectra discussed below, thus confirming the structural changes in the probed sample volume.

Detailed changes in local intensities in the XPD patterns are due to
the change in the lattice symmetry, {\it i.e.} the in-plane lattice expansion contrasts with the almost constant out-of-plane lattice parameter along the $c$-axis. The XPD patterns are element-selective in the sense that the presented patterns reveal the lattice structure from the viewpoint of the Eu(II) ions in our case.
For the case of EuIr$_2$Si$_2$, it has been shown that photoelectron diffraction for $4d$ - $4f$ resonant photoelectron excitation probes the Eu valency~\cite{Usachov2020}.
Since the lattice structure is known from X-ray diffraction, a detailed discussion of the XPD patterns is omitted here.

Figures~\ref{Fig4}(a-h) show the drastic change in the Fermi surface between 25~K and 300~K. At 25~K, the section through the Fermi surface in the highly symmetry $\Gamma$-$\Sigma$-X plane shows ellipses centered on the X-points as the brightest features with the long axis pointing along the $\Gamma-X$ direction. At room temperature, the ellipses have collapsed into elongated point features. 

The $E_B$ versus $k_x$ sections are dominated by two dispersion-free bands of very high intensity, representing the localized Eu  $4f$ and Pd $4d$ states at binding energies of 0.5~eV and 3~eV [see Fig.~\ref{Fig2}(c)]. 
To extract the dispersion of the bands intersecting the Fermi level, we normalized the measured intensity by the momentum-averaged energy distribution curve. This data processing reveals the band dispersion of the conduction bands [see Fig.~\ref{Fig4}(i-p)]. 
At 25~K, the corresponding intensity distribution shows the electron bands near the X-points with an electron-like parabolic dispersion [see Fig.~\ref{Fig4}(i)]. The effective mass of this band is $m/m_0=1$. At 300~K this band has shifted to a higher energy and thus becomes almost unoccupied [see Fig.~\ref{Fig2}(m)]. As this change is accompanied by the valence change from Eu(III) to Eu(II), it is likely that the electron occupying the free-electron-like state at low temperature has become localized at 300~K, explaining the increase in the peak at $E_B = 0.5$~eV and the decrease in conductivity observed in previous publications. Further details of the band dispersion characteristics are
discussed below in comparison with {\it ab-initio} DFT calculations.

\section{Theoretical Results}

\begin{figure*}
\includegraphics[width=0.9\textwidth]{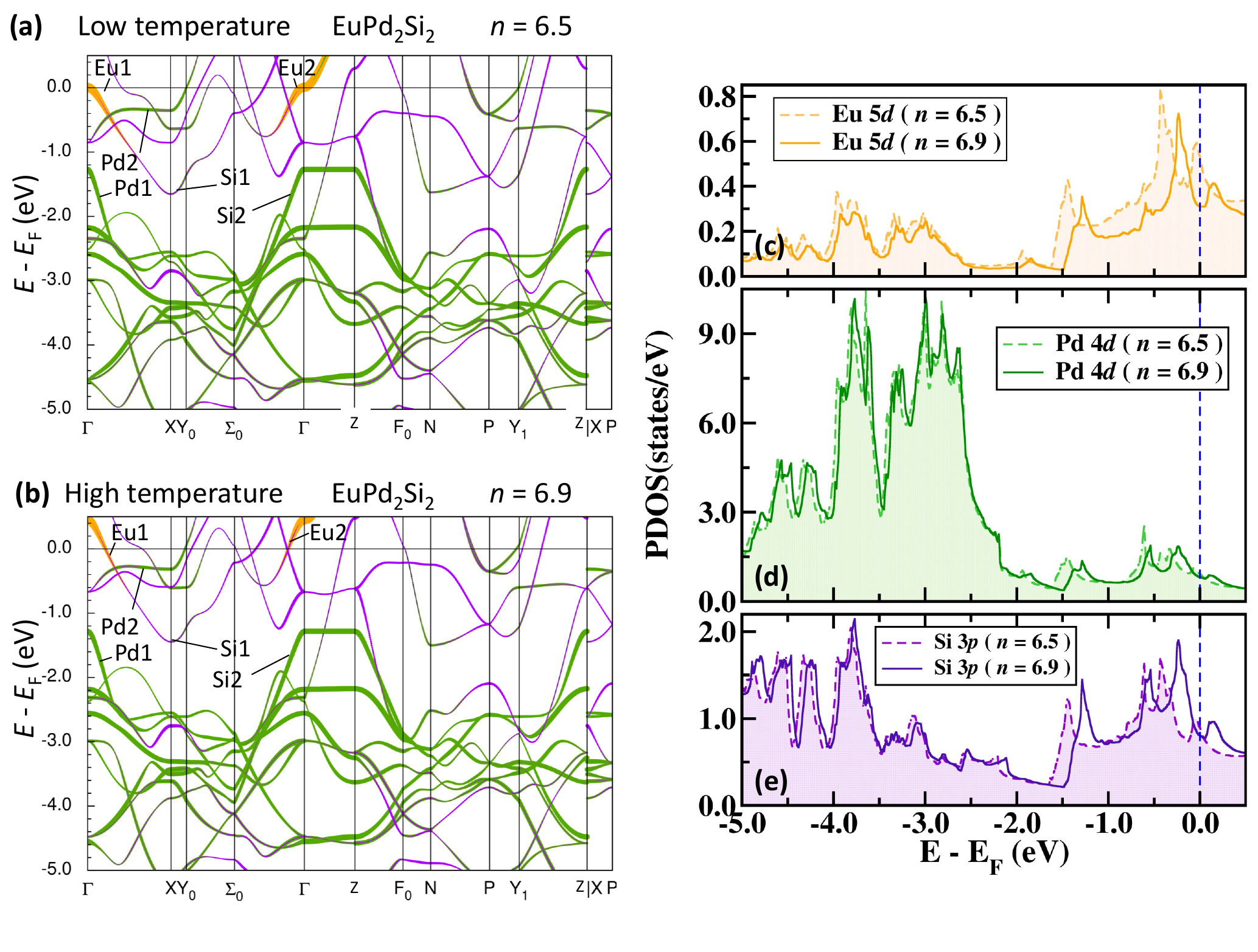}
\caption{\label{Fig5} 
Full-potential local-orbital calculations of the 
EuPd$_2$Si$_2$ compound for high symmetry directions in momentum space for valence numbers $n=6.5$ (a) and $n=6.9$ (b), corresponding to the low (a) and high (b) temperature phases. 
For the definition of the high symmetry points see 
Fig.~\ref{Fig1}(b).
Orbital projections shown in the fat band representation.
Orange bands correspond to Eu $5d$-derived bands, 
green to Pd $4d$-derived bands, and purple to Si $3p$-derived bands. 
Orbital-projected density-of-states functions (PDOS) for $n=6.5$ and $n=6.9$  for (c) Eu $5d$, (d) Pd $4d$, and (e) Si $3p$.
}
\end{figure*}

All density functional theory (DFT) calculations were carried out based on the full-potential local-orbital minimum-basis as implemented in FPLO~\cite{Koepernik1999,Opahle1999} employing local density approximation (LDA) for the exchange-correlation functional. 
Crystal information of the tetragonal bulk EuPd$_2$Si$_2$ (I4/mmm) as determined experimentally~\cite{Kliemt2022,Mitsuda2000,Palenzona1987} was adopted for the calculations. The first Brillouin zone was sampled by a 21$\times$21$\times$21 $k$-mesh. In all calculations, the Eu $4f$ orbitals were treated as core states while fixing the $4f$ occupancy ($n$) to the values determined here by the XPS evaluation using the open-core approximation as implemented in FPLO. 
As a consequence, the Eu $4f$ states are not shown in the electronic structure results. 
Figures~\ref{Fig5}(a,b) show LDA orbital-projected band structures of EuPd$2$Si$_2$ along the high symmetry paths as shown in Fig.~\ref{Fig1} (b) where each orbital contribution is colored in orange (Eu $5d$), green (Pd $4d$), and purple (Si $3p$).

The number of valence electrons was set to $n= 6.5$ for the low temperature structure 
and to $n=6.9$ for the high temperature structure.
These values correspond to the change of the mixed Eu(II)/Eu(III) valence as determined by the XPD evaluation shown in Fig.~\ref{Fig2}.
These Eu-derived bands (Eu1 and Eu2, Fig.~\ref{Fig5}) shift to a lower binding energy by about 0.5~eV
when the valence number is changed from $n= 6.5$ to 6.9, corresponding to the high temperature phase. 
The Eu $5d$ bands (orange) are thus less occupied at higher temperatures. 

Similarly, bands originating from Si orbitals (purple, band Si1) and Pd $4d$ orbitals (green, band Pd1) shift to lower binding energy with increasing valence number $n$. 
In contrast, at different momentum values the Pd $4d$ bands (green, band Pd$_b$) are only marginally changed by the different occupation number $n$.

The band shifts are also seen in the density-of-states functions projected onto the Eu $5d$, Pd $4d$, and Si $3p$ bands as shown in Figs.~\ref{Fig5}(c,d,e).
Here, the Pd $4d$ bands (green) form the largest contribution to the density of states for $E_B= 3$ to 4~eV. The center of mass of this large peak is shifted to lower binding energies with the increased $4f$ occupation number $n$, \textit{i.e.} in the same direction as the shift of the Eu $5d$ and Si $3p$ bands.
This is in good agreement with the experimental data shown in Fig.~\ref{Fig2}(c).

\section{Discussion}

\begin{figure}
\includegraphics[width=\columnwidth]{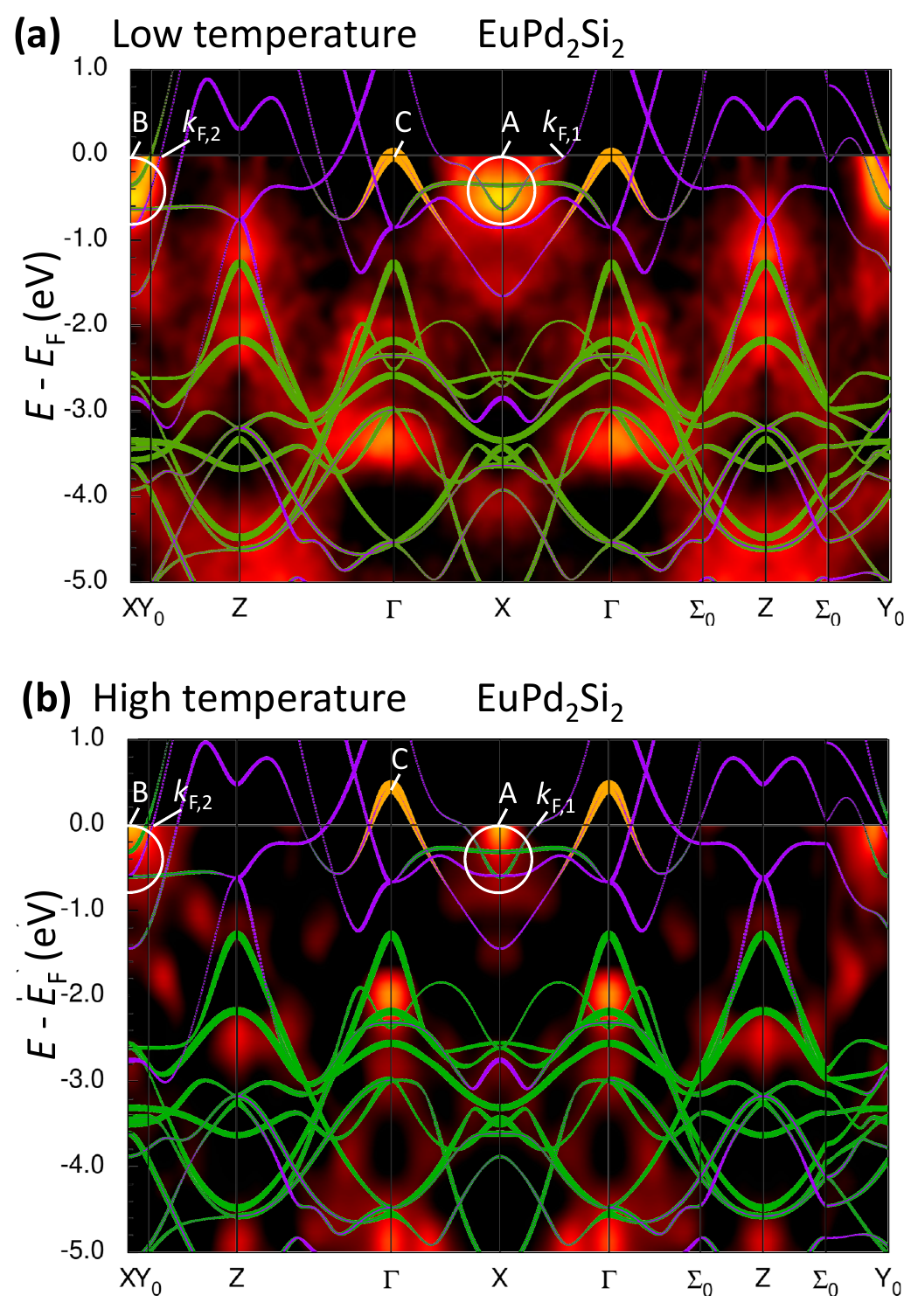}
\caption{\label{Fig6} 
Experimental photoemission intensity sections along the measured high symmetry directions for the (a) low and (b) high temperature data (Same color scales as in Fig.~\ref{Fig4}). Overlay of orbital-projected bands from 
full-potential local-orbital calculations of the 
EuPd$_2$Si$_2$ compound for selected high symmetry directions in momentum space for valence numbers $n=6.5$ (a) and $n=6.9$ (b), corresponding to the low (a) and high (b) temperature phases.
}
\end{figure}

We directly compare the calculated band structure with the experimentally obtained $E_B$ {\it versus} momentum $k$ sections
using the data overlay shown in Figs.~\ref{Fig6}(a,b).
The calculated electron-like  band with the minimum at the X-point at $E_B=0.6$~eV (areas A and B  in Fig.~\ref{Fig6}(a)) and the Fermi vector (mark $k_{F,1}$) between the $\Gamma$- and the X-point and the Fermi vector close to the Y-point (area B), fits quite well to the experimental electron-like band with high intensity. 
The band maxima at the $\Gamma$- point near the Fermi level  
[C in  Figs.~\ref{Fig6}(a)] has a pronounced Eu $5d$ orbital character.
Due to the decreasing photoemission cross section with increasing orbital number it does not show up in the experimental data.
The weaker photoemission intensities at a binding energy between $E_B=2$ and 3~eV correspond to the calculated Pd $4d$ bands.

At the Z-point, the photoemission data show a weak electron-like band with a minimum at $E_B=0.8$~eV. 
This band corresponds to the calculated band in the same energy range of hybridized Si $3p$ - Pd $4d$ character.

Now, we turn to the high-temperature (300~K) data. The high-temperature data show a lower signal-to-background ratio due to the significantly increased Debye-Waller scattering of photoemitted electrons, resulting in an increase of detected electrons after scattering events at the expense of electrons from direct photoemission processes. 
The intensity of scattered electrons is less dominant at the Fermi level. Here, we observe a shift 
of the $E_B=0.6$~eV minimum  at the X-point (A)
towards the Fermi level, combined with a decrease of the Fermi wave vector of this electron-like band ($k_{F,1}$ and $k_{F,2}$). 
In our calculations we observe  a similar band shift.



The calculated bands with at binding energies between 
$E_B=2$ and 3~eV fit to the experimental data. 
The theoretical and experimental bands show a shift to lower binding energies compared to the low-temperature data.
This shift corresponds to the shift of the density-of-states maximum
in the same binding energy range in Fig.~\ref{Fig2}(c) and to the calculated results shown in Fig.~\ref{Fig5}(c,d,e).



\begin{figure}
\includegraphics[width=\columnwidth]{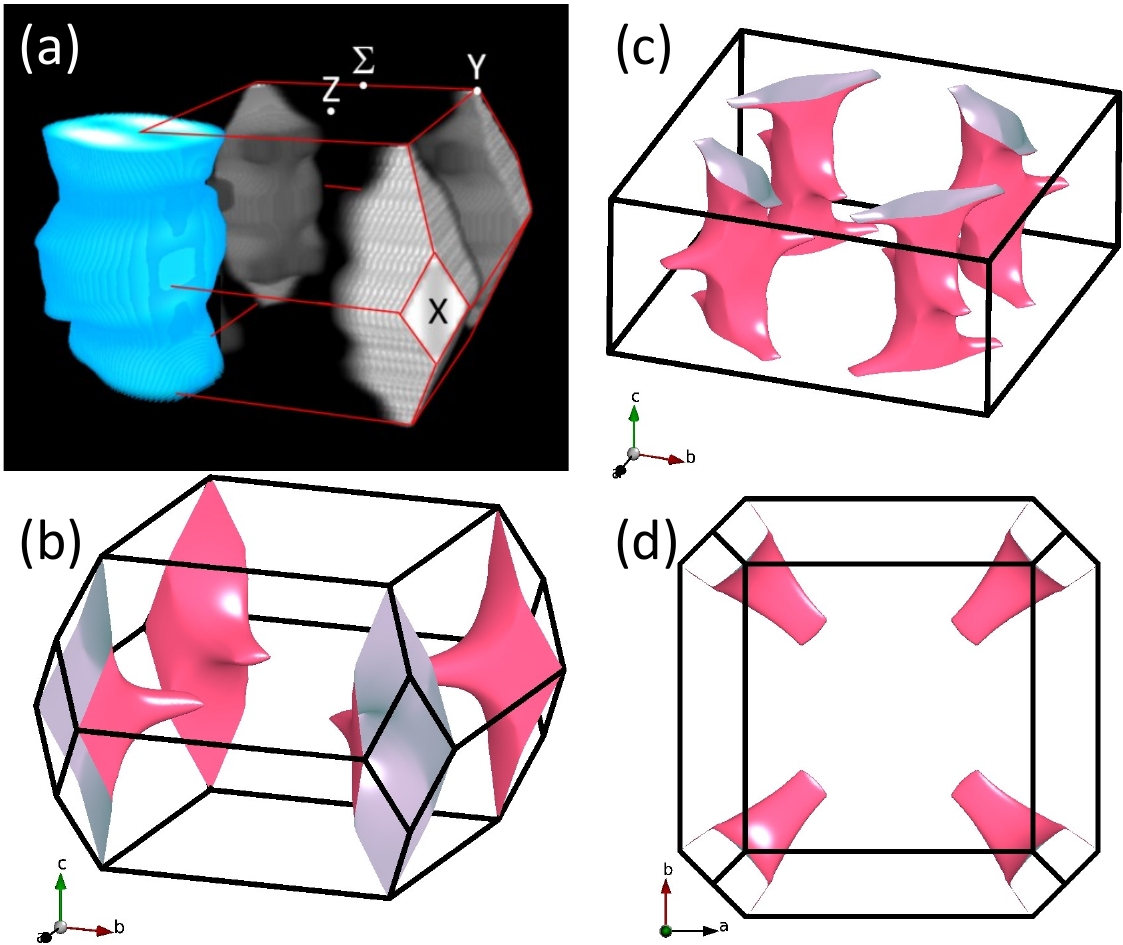}
\caption{\label{Fig7} 
(a) Experimentally determined Fermi surface (grey) within the Brillouin zone (red lines). Blue color shows one out of four Fermi surface sheet in the extended Brillouin zone scheme. (b) Calculated Fermi surface.
(c) Calculated Fermi surface in the extended Brillouin zone scheme. (d) Top view of the data shown in (b).
}
\end{figure}

For further comparison of experiment and theory, we show in Fig.~\ref{Fig7} the Fermi surface derived from the photoemission intensity and from the calculated band structure.
Taking advantage of the fact that the free electron-like final states are located on a spherical surface, we determined the Fermi surface at different cuts through the three-dimensional Brillouin zone parallel to the $\Gamma$-$\Sigma$-X plane but at different $k_z$ values~\cite{Agustsson2021}.
These slices are then concatenated to the Fermi surface in the three-dimensional Brillouin zone. 
The three-dimensional Fermi surface in the Brillouin zone shown in Fig.~\ref{Fig7}(a) for the low temperature data shows the Fermi surface sheets at the edges of the Brillouin zone near the X-points. 
We used blue for designating the Fermi surface extended  into the adjacent Brillouin zones. This Fermi surface sheet reveals the tubular structure of this sheet, indicating the almost two-dimensional character of this band. In direct space, the corresponding electron state is predominantly localized within the plane containing the Si ions [Fig.~\ref{Fig1}(a)].

The Fermi surface derived from the density functional calculation  [Fig.~\ref{Fig7}(b-d)] shows a shape of the Fermi surface similar to the experimental one. To emphasize the tubular character of the Fermi surface sheet, the theoretical data are also shown in an extended Brillouin zone scheme
[Fig.~\ref{Fig7}(c)]. Even the details of the elliptical cross sections with the long axis along $\Gamma$ - X and the short axis along Z - $\Sigma$ appear similar to the experimental ones.

\hspace{1cm}

\section{Conclusion}

We have experimentally studied the microscopic mechanism of the valence
transition of Eu ions in EuPd$_2$Si$_2$.
We determined the valence transition of the Eu ions using hard X-ray photoelectron spectroscopy, resulting
in an Eu ion charge of $v=2.1$ at high temperature and $v=2.5$ at low temperature, corresponding to an Eu $4f$ occupation number $n=6.9$ and
6.5, respectively.
By evaluating the hard X-ray photoemission diffraction patterns, we confirm the volume contraction at low temperatures in the probed sample volume.
Using hard X-ray angle-resolved photoemission spectroscopy, 
we studied the evolution of the valence transition in the bulk band structure.
We observed a pronounced shrinkage of a Fermi surface sheet from low temperature to room temperature. This is due to the 
up-shift of a hybridized Si $3p$-Pd $4d$ valence band that is hybridized with Eu $5d$ orbitals (see $k_{F,1}$ in Fig.~\ref{Fig7}(a,b)). 
The up-shift in energy is associated with a decrease in the occupation number of conduction band states and to an increase in the 
Eu $4f$ occupation number.
This observation suggests a partial charge transfer between Eu and Pd/Si sites. 

The combined experimental and theoretical effort allows to understand the changes of the momentum-resolved electronic structure in this compound. 
The valence transition causes a charge transfer from localized Eu $4f$ states to conduction band states at low temperature. The most significant changes are observed for those bands that exhibit a partial Eu $5d$ orbital character.

\acknowledgements
This work was funded by the Deutsche Forschungsgemeinschaft (DFG, German Research Foundation), grant no. TRR288–422213477 (projects B04, A03, and A05),
and by the BMBF (projects 05K22UM2 and 05K22UM4). 
Funding for the instrument by the Federal Ministry of Education and Research (BMBF) under framework program ErUM is gratefully acknowledged.
We acknowledge DESY (Hamburg, Germany), a member of the Helmholtz Association HGF, for the provision of experimental facilities. Parts of this research were carried out at PETRA III using beamline P22.  
O.F. acknowledges funding by TopDyn.
H.J.E. thanks Denys Vyalikh for fruitful discussions.


\begin{thebibliography}{0}%
\makeatletter
\providecommand \@ifxundefined [1]{%
 \@ifx{#1\undefined}
}%
\providecommand \@ifnum [1]{%
 \ifnum #1\expandafter \@firstoftwo
 \else \expandafter \@secondoftwo
 \fi
}%
\providecommand \@ifx [1]{%
 \ifx #1\expandafter \@firstoftwo
 \else \expandafter \@secondoftwo
 \fi
}%
\providecommand \natexlab [1]{#1}%
\providecommand \enquote  [1]{``#1''}%
\providecommand \bibnamefont  [1]{#1}%
\providecommand \bibfnamefont [1]{#1}%
\providecommand \citenamefont [1]{#1}%
\providecommand \href@noop [0]{\@secondoftwo}%
\providecommand \href [0]{\begingroup \@sanitize@url \@href}%
\providecommand \@href[1]{\@@startlink{#1}\@@href}%
\providecommand \@@href[1]{\endgroup#1\@@endlink}%
\providecommand \@sanitize@url [0]{\catcode `\\12\catcode `\$12\catcode
  `\&12\catcode `\#12\catcode `\^12\catcode `\_12\catcode `\%12\relax}%
\providecommand \@@startlink[1]{}%
\providecommand \@@endlink[0]{}%
\providecommand \url  [0]{\begingroup\@sanitize@url \@url }%
\providecommand \@url [1]{\endgroup\@href {#1}{\urlprefix }}%
\providecommand \urlprefix  [0]{URL }%
\providecommand \Eprint [0]{\href }%
\providecommand \doibase [0]{http://dx.doi.org/}%
\providecommand \selectlanguage [0]{\@gobble}%
\providecommand \bibinfo  [0]{\@secondoftwo}%
\providecommand \bibfield  [0]{\@secondoftwo}%
\providecommand \translation [1]{[#1]}%
\providecommand \BibitemOpen [0]{}%
\providecommand \bibitemStop [0]{}%
\providecommand \bibitemNoStop [0]{.\EOS\space}%
\providecommand \EOS [0]{\spacefactor3000\relax}%
\providecommand \BibitemShut  [1]{\csname bibitem#1\endcsname}%
\let\auto@bib@innerbib\@empty
\end{thebibliography}%


\begin{thebibliography}{42}%
\makeatletter
\providecommand \@ifxundefined [1]{%
 \@ifx{#1\undefined}
}%
\providecommand \@ifnum [1]{%
 \ifnum #1\expandafter \@firstoftwo
 \else \expandafter \@secondoftwo
 \fi
}%
\providecommand \@ifx [1]{%
 \ifx #1\expandafter \@firstoftwo
 \else \expandafter \@secondoftwo
 \fi
}%
\providecommand \natexlab [1]{#1}%
\providecommand \enquote  [1]{``#1''}%
\providecommand \bibnamefont  [1]{#1}%
\providecommand \bibfnamefont [1]{#1}%
\providecommand \citenamefont [1]{#1}%
\providecommand \href@noop [0]{\@secondoftwo}%
\providecommand \href [0]{\begingroup \@sanitize@url \@href}%
\providecommand \@href[1]{\@@startlink{#1}\@@href}%
\providecommand \@@href[1]{\endgroup#1\@@endlink}%
\providecommand \@sanitize@url [0]{\catcode `\\12\catcode `\$12\catcode
  `\&12\catcode `\#12\catcode `\^12\catcode `\_12\catcode `\%12\relax}%
\providecommand \@@startlink[1]{}%
\providecommand \@@endlink[0]{}%
\providecommand \url  [0]{\begingroup\@sanitize@url \@url }%
\providecommand \@url [1]{\endgroup\@href {#1}{\urlprefix }}%
\providecommand \urlprefix  [0]{URL }%
\providecommand \Eprint [0]{\href }%
\providecommand \doibase [0]{http://dx.doi.org/}%
\providecommand \selectlanguage [0]{\@gobble}%
\providecommand \bibinfo  [0]{\@secondoftwo}%
\providecommand \bibfield  [0]{\@secondoftwo}%
\providecommand \translation [1]{[#1]}%
\providecommand \BibitemOpen [0]{}%
\providecommand \bibitemStop [0]{}%
\providecommand \bibitemNoStop [0]{.\EOS\space}%
\providecommand \EOS [0]{\spacefactor3000\relax}%
\providecommand \BibitemShut  [1]{\csname bibitem#1\endcsname}%
\let\auto@bib@innerbib\@empty
\bibitem [{\citenamefont {Wirth}\ and\ \citenamefont
  {Steglich}(2016)}]{Wirth2016}%
  \BibitemOpen
  \bibfield  {author} {\bibinfo {author} {\bibfnamefont {S.}~\bibnamefont
  {Wirth}}\ and\ \bibinfo {author} {\bibfnamefont {F.}~\bibnamefont
  {Steglich}},\ }\href {\doibase 10.1038/natrevmats.2016.51} {\bibfield
  {journal} {\bibinfo  {journal} {Nature Reviews Materials}\ }\textbf {\bibinfo
  {volume} {1}} (\bibinfo {year} {2016}),\
  10.1038/natrevmats.2016.51}\BibitemShut {NoStop}%
\bibitem [{\citenamefont {Rau}\ and\ \citenamefont {Gingras}(2019)}]{Rau2019}%
  \BibitemOpen
  \bibfield  {author} {\bibinfo {author} {\bibfnamefont {J.~G.}\ \bibnamefont
  {Rau}}\ and\ \bibinfo {author} {\bibfnamefont {M.~J.}\ \bibnamefont
  {Gingras}},\ }\href {\doibase 10.1146/annurev-conmatphys-022317-110520}
  {\bibfield  {journal} {\bibinfo  {journal} {Annual Review of Condensed Matter
  Physics}\ }\textbf {\bibinfo {volume} {10}},\ \bibinfo {pages} {357}
  (\bibinfo {year} {2019})}\BibitemShut {NoStop}%
\bibitem [{\citenamefont {Pfleiderer}(2009)}]{Pfleiderer2009}%
  \BibitemOpen
  \bibfield  {author} {\bibinfo {author} {\bibfnamefont {C.}~\bibnamefont
  {Pfleiderer}},\ }\href {\doibase 10.1103/revmodphys.81.1551} {\bibfield
  {journal} {\bibinfo  {journal} {Reviews of Modern Physics}\ }\textbf
  {\bibinfo {volume} {81}},\ \bibinfo {pages} {1551} (\bibinfo {year}
  {2009})}\BibitemShut {NoStop}%
\bibitem [{\citenamefont {Stewart}(2001)}]{Stewart2001}%
  \BibitemOpen
  \bibfield  {author} {\bibinfo {author} {\bibfnamefont {G.~R.}\ \bibnamefont
  {Stewart}},\ }\href {\doibase 10.1103/revmodphys.73.797} {\bibfield
  {journal} {\bibinfo  {journal} {Reviews of Modern Physics}\ }\textbf
  {\bibinfo {volume} {73}},\ \bibinfo {pages} {797} (\bibinfo {year}
  {2001})}\BibitemShut {NoStop}%
\bibitem [{\citenamefont {{\={O}}nuki}\ \emph {et~al.}(2020)\citenamefont
  {{\={O}}nuki}, \citenamefont {Hedo},\ and\ \citenamefont
  {Honda}}]{Onuki2020}%
  \BibitemOpen
  \bibfield  {author} {\bibinfo {author} {\bibfnamefont {Y.}~\bibnamefont
  {{\={O}}nuki}}, \bibinfo {author} {\bibfnamefont {M.}~\bibnamefont {Hedo}}, \
  and\ \bibinfo {author} {\bibfnamefont {F.}~\bibnamefont {Honda}},\ }\href
  {\doibase 10.7566/jpsj.89.102001} {\bibfield  {journal} {\bibinfo  {journal}
  {Journal of the Physical Society of Japan}\ }\textbf {\bibinfo {volume}
  {89}},\ \bibinfo {pages} {102001} (\bibinfo {year} {2020})}\BibitemShut
  {NoStop}%
\bibitem [{\citenamefont {{\={O}}nuki}\ \emph {et~al.}(2016)\citenamefont
  {{\={O}}nuki}, \citenamefont {Nakamura}, \citenamefont {Honda}, \citenamefont
  {Aoki}, \citenamefont {Tekeuchi}, \citenamefont {Nakashima}, \citenamefont
  {Amako}, \citenamefont {Harima}, \citenamefont {Matsubayashi}, \citenamefont
  {Uwatoko}, \citenamefont {Kayama}, \citenamefont {Kagayama}, \citenamefont
  {Shimizu}, \citenamefont {Muthu}, \citenamefont {Braithwaite}, \citenamefont
  {Salce}, \citenamefont {Shiba}, \citenamefont {Yara}, \citenamefont
  {Ashitomi}, \citenamefont {Akamine}, \citenamefont {Tomori}, \citenamefont
  {Hedo},\ and\ \citenamefont {Nakama}}]{Onuki2016}%
  \BibitemOpen
  \bibfield  {author} {\bibinfo {author} {\bibfnamefont {Y.}~\bibnamefont
  {{\={O}}nuki}}, \bibinfo {author} {\bibfnamefont {A.}~\bibnamefont
  {Nakamura}}, \bibinfo {author} {\bibfnamefont {F.}~\bibnamefont {Honda}},
  \bibinfo {author} {\bibfnamefont {D.}~\bibnamefont {Aoki}}, \bibinfo {author}
  {\bibfnamefont {T.}~\bibnamefont {Tekeuchi}}, \bibinfo {author}
  {\bibfnamefont {M.}~\bibnamefont {Nakashima}}, \bibinfo {author}
  {\bibfnamefont {Y.}~\bibnamefont {Amako}}, \bibinfo {author} {\bibfnamefont
  {H.}~\bibnamefont {Harima}}, \bibinfo {author} {\bibfnamefont
  {K.}~\bibnamefont {Matsubayashi}}, \bibinfo {author} {\bibfnamefont
  {Y.}~\bibnamefont {Uwatoko}}, \bibinfo {author} {\bibfnamefont
  {S.}~\bibnamefont {Kayama}}, \bibinfo {author} {\bibfnamefont
  {T.}~\bibnamefont {Kagayama}}, \bibinfo {author} {\bibfnamefont
  {K.}~\bibnamefont {Shimizu}}, \bibinfo {author} {\bibfnamefont {S.~E.}\
  \bibnamefont {Muthu}}, \bibinfo {author} {\bibfnamefont {D.}~\bibnamefont
  {Braithwaite}}, \bibinfo {author} {\bibfnamefont {B.}~\bibnamefont {Salce}},
  \bibinfo {author} {\bibfnamefont {H.}~\bibnamefont {Shiba}}, \bibinfo
  {author} {\bibfnamefont {T.}~\bibnamefont {Yara}}, \bibinfo {author}
  {\bibfnamefont {Y.}~\bibnamefont {Ashitomi}}, \bibinfo {author}
  {\bibfnamefont {H.}~\bibnamefont {Akamine}}, \bibinfo {author} {\bibfnamefont
  {K.}~\bibnamefont {Tomori}}, \bibinfo {author} {\bibfnamefont
  {M.}~\bibnamefont {Hedo}}, \ and\ \bibinfo {author} {\bibfnamefont
  {T.}~\bibnamefont {Nakama}},\ }\href {\doibase 10.1080/14786435.2016.1218081}
  {\bibfield  {journal} {\bibinfo  {journal} {Philosophical Magazine}\ }\textbf
  {\bibinfo {volume} {97}},\ \bibinfo {pages} {3399} (\bibinfo {year}
  {2016})}\BibitemShut {NoStop}%
\bibitem [{\citenamefont {Xu}\ \emph {et~al.}(2020)\citenamefont {Xu},
  \citenamefont {Elcoro}, \citenamefont {Song}, \citenamefont {Wieder},
  \citenamefont {Vergniory}, \citenamefont {Regnault}, \citenamefont {Chen},
  \citenamefont {Felser},\ and\ \citenamefont {Bernevig}}]{Xu2020}%
  \BibitemOpen
  \bibfield  {author} {\bibinfo {author} {\bibfnamefont {Y.}~\bibnamefont
  {Xu}}, \bibinfo {author} {\bibfnamefont {L.}~\bibnamefont {Elcoro}}, \bibinfo
  {author} {\bibfnamefont {Z.-D.}\ \bibnamefont {Song}}, \bibinfo {author}
  {\bibfnamefont {B.~J.}\ \bibnamefont {Wieder}}, \bibinfo {author}
  {\bibfnamefont {M.~G.}\ \bibnamefont {Vergniory}}, \bibinfo {author}
  {\bibfnamefont {N.}~\bibnamefont {Regnault}}, \bibinfo {author}
  {\bibfnamefont {Y.}~\bibnamefont {Chen}}, \bibinfo {author} {\bibfnamefont
  {C.}~\bibnamefont {Felser}}, \ and\ \bibinfo {author} {\bibfnamefont {B.~A.}\
  \bibnamefont {Bernevig}},\ }\href {\doibase 10.1038/s41586-020-2837-0}
  {\bibfield  {journal} {\bibinfo  {journal} {Nature}\ }\textbf {\bibinfo
  {volume} {586}},\ \bibinfo {pages} {702} (\bibinfo {year}
  {2020})}\BibitemShut {NoStop}%
\bibitem [{\citenamefont {Rahn}\ \emph {et~al.}(2018)\citenamefont {Rahn},
  \citenamefont {Soh}, \citenamefont {Francoual}, \citenamefont {Veiga},
  \citenamefont {Strempfer}, \citenamefont {Mardegan}, \citenamefont {Yan},
  \citenamefont {Guo}, \citenamefont {Shi},\ and\ \citenamefont
  {Boothroyd}}]{Rahn2018}%
  \BibitemOpen
  \bibfield  {author} {\bibinfo {author} {\bibfnamefont {M.~C.}\ \bibnamefont
  {Rahn}}, \bibinfo {author} {\bibfnamefont {J.-R.}\ \bibnamefont {Soh}},
  \bibinfo {author} {\bibfnamefont {S.}~\bibnamefont {Francoual}}, \bibinfo
  {author} {\bibfnamefont {L.~S.~I.}\ \bibnamefont {Veiga}}, \bibinfo {author}
  {\bibfnamefont {J.}~\bibnamefont {Strempfer}}, \bibinfo {author}
  {\bibfnamefont {J.}~\bibnamefont {Mardegan}}, \bibinfo {author}
  {\bibfnamefont {D.~Y.}\ \bibnamefont {Yan}}, \bibinfo {author} {\bibfnamefont
  {Y.~F.}\ \bibnamefont {Guo}}, \bibinfo {author} {\bibfnamefont {Y.~G.}\
  \bibnamefont {Shi}}, \ and\ \bibinfo {author} {\bibfnamefont {A.~T.}\
  \bibnamefont {Boothroyd}},\ }\href {\doibase 10.1103/physrevb.97.214422}
  {\bibfield  {journal} {\bibinfo  {journal} {Physical Review B}\ }\textbf
  {\bibinfo {volume} {97}},\ \bibinfo {pages} {214422} (\bibinfo {year}
  {2018})}\BibitemShut {NoStop}%
\bibitem [{\citenamefont {Schulz}\ \emph {et~al.}(2019)\citenamefont {Schulz},
  \citenamefont {Nechaev}, \citenamefont {Güttler}, \citenamefont {Poelchen},
  \citenamefont {Generalov}, \citenamefont {Danzenbächer}, \citenamefont
  {Chikina}, \citenamefont {Seiro}, \citenamefont {Kliemt}, \citenamefont
  {Vyazovskaya}, \citenamefont {Kim}, \citenamefont {Dudin}, \citenamefont
  {Chulkov}, \citenamefont {Laubschat}, \citenamefont {Krasovskii},
  \citenamefont {Geibel}, \citenamefont {Krellner}, \citenamefont {Kummer},\
  and\ \citenamefont {Vyalikh}}]{Schulz2019}%
  \BibitemOpen
  \bibfield  {author} {\bibinfo {author} {\bibfnamefont {S.}~\bibnamefont
  {Schulz}}, \bibinfo {author} {\bibfnamefont {I.~A.}\ \bibnamefont {Nechaev}},
  \bibinfo {author} {\bibfnamefont {M.}~\bibnamefont {Güttler}}, \bibinfo
  {author} {\bibfnamefont {G.}~\bibnamefont {Poelchen}}, \bibinfo {author}
  {\bibfnamefont {A.}~\bibnamefont {Generalov}}, \bibinfo {author}
  {\bibfnamefont {S.}~\bibnamefont {Danzenbächer}}, \bibinfo {author}
  {\bibfnamefont {A.}~\bibnamefont {Chikina}}, \bibinfo {author} {\bibfnamefont
  {S.}~\bibnamefont {Seiro}}, \bibinfo {author} {\bibfnamefont
  {K.}~\bibnamefont {Kliemt}}, \bibinfo {author} {\bibfnamefont {A.~Y.}\
  \bibnamefont {Vyazovskaya}}, \bibinfo {author} {\bibfnamefont {T.~K.}\
  \bibnamefont {Kim}}, \bibinfo {author} {\bibfnamefont {P.}~\bibnamefont
  {Dudin}}, \bibinfo {author} {\bibfnamefont {E.~V.}\ \bibnamefont {Chulkov}},
  \bibinfo {author} {\bibfnamefont {C.}~\bibnamefont {Laubschat}}, \bibinfo
  {author} {\bibfnamefont {E.~E.}\ \bibnamefont {Krasovskii}}, \bibinfo
  {author} {\bibfnamefont {C.}~\bibnamefont {Geibel}}, \bibinfo {author}
  {\bibfnamefont {C.}~\bibnamefont {Krellner}}, \bibinfo {author}
  {\bibfnamefont {K.}~\bibnamefont {Kummer}}, \ and\ \bibinfo {author}
  {\bibfnamefont {D.~V.}\ \bibnamefont {Vyalikh}},\ }\href {\doibase
  10.1038/s41535-019-0166-z} {\bibfield  {journal} {\bibinfo  {journal} {npj
  Quantum Materials}\ }\textbf {\bibinfo {volume} {4}} (\bibinfo {year}
  {2019}),\ 10.1038/s41535-019-0166-z}\BibitemShut {NoStop}%
\bibitem [{\citenamefont {Sampathkumaran}\ \emph {et~al.}(1981)\citenamefont
  {Sampathkumaran}, \citenamefont {Gupta}, \citenamefont {Vijayaraghavan},
  \citenamefont {Gopalakrishnan}, \citenamefont {Pillay},\ and\ \citenamefont
  {Devare}}]{Sampathkumaran1981}%
  \BibitemOpen
  \bibfield  {author} {\bibinfo {author} {\bibfnamefont {E.~V.}\ \bibnamefont
  {Sampathkumaran}}, \bibinfo {author} {\bibfnamefont {L.~C.}\ \bibnamefont
  {Gupta}}, \bibinfo {author} {\bibfnamefont {R.}~\bibnamefont
  {Vijayaraghavan}}, \bibinfo {author} {\bibfnamefont {K.~V.}\ \bibnamefont
  {Gopalakrishnan}}, \bibinfo {author} {\bibfnamefont {R.~G.}\ \bibnamefont
  {Pillay}}, \ and\ \bibinfo {author} {\bibfnamefont {H.~G.}\ \bibnamefont
  {Devare}},\ }\href {\doibase 10.1088/0022-3719/14/9/006} {\bibfield
  {journal} {\bibinfo  {journal} {Journal of Physics C: Solid State Physics}\
  }\textbf {\bibinfo {volume} {14}},\ \bibinfo {pages} {L237} (\bibinfo {year}
  {1981})}\BibitemShut {NoStop}%
\bibitem [{\citenamefont {Bauminger}\ \emph {et~al.}(1973)\citenamefont
  {Bauminger}, \citenamefont {Froindlich}, \citenamefont {Nowik}, \citenamefont
  {Ofer}, \citenamefont {Felner},\ and\ \citenamefont {Mayer}}]{Bauminger1973}%
  \BibitemOpen
  \bibfield  {author} {\bibinfo {author} {\bibfnamefont {E.~R.}\ \bibnamefont
  {Bauminger}}, \bibinfo {author} {\bibfnamefont {D.}~\bibnamefont
  {Froindlich}}, \bibinfo {author} {\bibfnamefont {I.}~\bibnamefont {Nowik}},
  \bibinfo {author} {\bibfnamefont {S.}~\bibnamefont {Ofer}}, \bibinfo {author}
  {\bibfnamefont {I.}~\bibnamefont {Felner}}, \ and\ \bibinfo {author}
  {\bibfnamefont {I.}~\bibnamefont {Mayer}},\ }\href {\doibase
  10.1103/physrevlett.30.1053} {\bibfield  {journal} {\bibinfo  {journal}
  {Physical Review Letters}\ }\textbf {\bibinfo {volume} {30}},\ \bibinfo
  {pages} {1053} (\bibinfo {year} {1973})}\BibitemShut {NoStop}%
\bibitem [{\citenamefont {Patil}\ \emph {et~al.}(1993)\citenamefont {Patil},
  \citenamefont {Nagarajan}, \citenamefont {Godart}, \citenamefont {Kappler},
  \citenamefont {Gupta}, \citenamefont {Padalia},\ and\ \citenamefont
  {Vijayaraghavan}}]{Patil1993}%
  \BibitemOpen
  \bibfield  {author} {\bibinfo {author} {\bibfnamefont {S.}~\bibnamefont
  {Patil}}, \bibinfo {author} {\bibfnamefont {R.}~\bibnamefont {Nagarajan}},
  \bibinfo {author} {\bibfnamefont {C.}~\bibnamefont {Godart}}, \bibinfo
  {author} {\bibfnamefont {J.~P.}\ \bibnamefont {Kappler}}, \bibinfo {author}
  {\bibfnamefont {L.~C.}\ \bibnamefont {Gupta}}, \bibinfo {author}
  {\bibfnamefont {B.~D.}\ \bibnamefont {Padalia}}, \ and\ \bibinfo {author}
  {\bibfnamefont {R.}~\bibnamefont {Vijayaraghavan}},\ }\href {\doibase
  10.1103/physrevb.47.8794} {\bibfield  {journal} {\bibinfo  {journal}
  {Physical Review B}\ }\textbf {\bibinfo {volume} {47}},\ \bibinfo {pages}
  {8794} (\bibinfo {year} {1993})}\BibitemShut {NoStop}%
\bibitem [{\citenamefont {Chevalier}\ \emph {et~al.}(1986)\citenamefont
  {Chevalier}, \citenamefont {Coey}, \citenamefont {Lloret},\ and\
  \citenamefont {Etourneau}}]{Chevalier1986}%
  \BibitemOpen
  \bibfield  {author} {\bibinfo {author} {\bibfnamefont {B.}~\bibnamefont
  {Chevalier}}, \bibinfo {author} {\bibfnamefont {J.~M.~D.}\ \bibnamefont
  {Coey}}, \bibinfo {author} {\bibfnamefont {B.}~\bibnamefont {Lloret}}, \ and\
  \bibinfo {author} {\bibfnamefont {J.}~\bibnamefont {Etourneau}},\ }\href
  {\doibase 10.1088/0022-3719/19/23/015} {\bibfield  {journal} {\bibinfo
  {journal} {Journal of Physics C: Solid State Physics}\ }\textbf {\bibinfo
  {volume} {19}},\ \bibinfo {pages} {4521} (\bibinfo {year}
  {1986})}\BibitemShut {NoStop}%
\bibitem [{\citenamefont {Seiro}\ \emph {et~al.}(2019)\citenamefont {Seiro},
  \citenamefont {Prots}, \citenamefont {Kummer}, \citenamefont {Rosner},
  \citenamefont {Gil},\ and\ \citenamefont {Geibel}}]{Seiro2019}%
  \BibitemOpen
  \bibfield  {author} {\bibinfo {author} {\bibfnamefont {S.}~\bibnamefont
  {Seiro}}, \bibinfo {author} {\bibfnamefont {Y.}~\bibnamefont {Prots}},
  \bibinfo {author} {\bibfnamefont {K.}~\bibnamefont {Kummer}}, \bibinfo
  {author} {\bibfnamefont {H.}~\bibnamefont {Rosner}}, \bibinfo {author}
  {\bibfnamefont {R.~C.}\ \bibnamefont {Gil}}, \ and\ \bibinfo {author}
  {\bibfnamefont {C.}~\bibnamefont {Geibel}},\ }\href {\doibase
  10.1088/1361-648x/ab1509} {\bibfield  {journal} {\bibinfo  {journal} {Journal
  of Physics: Condensed Matter}\ }\textbf {\bibinfo {volume} {31}},\ \bibinfo
  {pages} {305602} (\bibinfo {year} {2019})}\BibitemShut {NoStop}%
\bibitem [{\citenamefont {Mitsuda}\ \emph {et~al.}(2012)\citenamefont
  {Mitsuda}, \citenamefont {Hamano}, \citenamefont {Araoka}, \citenamefont
  {Yayama},\ and\ \citenamefont {Wada}}]{Mitsuda2012}%
  \BibitemOpen
  \bibfield  {author} {\bibinfo {author} {\bibfnamefont {A.}~\bibnamefont
  {Mitsuda}}, \bibinfo {author} {\bibfnamefont {S.}~\bibnamefont {Hamano}},
  \bibinfo {author} {\bibfnamefont {N.}~\bibnamefont {Araoka}}, \bibinfo
  {author} {\bibfnamefont {H.}~\bibnamefont {Yayama}}, \ and\ \bibinfo {author}
  {\bibfnamefont {H.}~\bibnamefont {Wada}},\ }\href {\doibase
  10.1143/jpsj.81.023709} {\bibfield  {journal} {\bibinfo  {journal} {Journal
  of the Physical Society of Japan}\ }\textbf {\bibinfo {volume} {81}},\
  \bibinfo {pages} {023709} (\bibinfo {year} {2012})}\BibitemShut {NoStop}%
\bibitem [{\citenamefont {Nakamura}\ \emph {et~al.}(2012)\citenamefont
  {Nakamura}, \citenamefont {Nakama}, \citenamefont {Uchima}, \citenamefont
  {Arakaki}, \citenamefont {Zukeran}, \citenamefont {Komesu}, \citenamefont
  {Takeda}, \citenamefont {Takaesu}, \citenamefont {Nakamura}, \citenamefont
  {Hedo}, \citenamefont {Yagasaki},\ and\ \citenamefont
  {Uwatoko}}]{Nakamura2012}%
  \BibitemOpen
  \bibfield  {author} {\bibinfo {author} {\bibfnamefont {A.}~\bibnamefont
  {Nakamura}}, \bibinfo {author} {\bibfnamefont {T.}~\bibnamefont {Nakama}},
  \bibinfo {author} {\bibfnamefont {K.}~\bibnamefont {Uchima}}, \bibinfo
  {author} {\bibfnamefont {N.}~\bibnamefont {Arakaki}}, \bibinfo {author}
  {\bibfnamefont {C.}~\bibnamefont {Zukeran}}, \bibinfo {author} {\bibfnamefont
  {S.}~\bibnamefont {Komesu}}, \bibinfo {author} {\bibfnamefont
  {M.}~\bibnamefont {Takeda}}, \bibinfo {author} {\bibfnamefont
  {Y.}~\bibnamefont {Takaesu}}, \bibinfo {author} {\bibfnamefont
  {D.}~\bibnamefont {Nakamura}}, \bibinfo {author} {\bibfnamefont
  {M.}~\bibnamefont {Hedo}}, \bibinfo {author} {\bibfnamefont {K.}~\bibnamefont
  {Yagasaki}}, \ and\ \bibinfo {author} {\bibfnamefont {Y.}~\bibnamefont
  {Uwatoko}},\ }\href {\doibase 10.1088/1742-6596/400/3/032106} {\bibfield
  {journal} {\bibinfo  {journal} {Journal of Physics: Conference Series}\
  }\textbf {\bibinfo {volume} {400}},\ \bibinfo {pages} {032106} (\bibinfo
  {year} {2012})}\BibitemShut {NoStop}%
\bibitem [{\citenamefont {Dionicio}\ \emph {et~al.}(2006)\citenamefont
  {Dionicio}, \citenamefont {Wilhelm}, \citenamefont {Hossain},\ and\
  \citenamefont {Geibel}}]{Dionicio2006}%
  \BibitemOpen
  \bibfield  {author} {\bibinfo {author} {\bibfnamefont {G.}~\bibnamefont
  {Dionicio}}, \bibinfo {author} {\bibfnamefont {H.}~\bibnamefont {Wilhelm}},
  \bibinfo {author} {\bibfnamefont {Z.}~\bibnamefont {Hossain}}, \ and\
  \bibinfo {author} {\bibfnamefont {C.}~\bibnamefont {Geibel}},\ }\href
  {\doibase 10.1016/j.physb.2006.01.258} {\bibfield  {journal} {\bibinfo
  {journal} {Physica B: Condensed Matter}\ }\textbf {\bibinfo {volume}
  {378-380}},\ \bibinfo {pages} {724} (\bibinfo {year} {2006})}\BibitemShut
  {NoStop}%
\bibitem [{\citenamefont {Ichiki}\ \emph {et~al.}(2017)\citenamefont {Ichiki},
  \citenamefont {Mimura}, \citenamefont {Anzai}, \citenamefont {Uozumi},
  \citenamefont {Sato}, \citenamefont {Utsumi}, \citenamefont {Ueda},
  \citenamefont {Mitsuda}, \citenamefont {Wada}, \citenamefont {Taguchi},
  \citenamefont {Shimada}, \citenamefont {Namatame},\ and\ \citenamefont
  {Taniguchi}}]{Ichiki2017}%
  \BibitemOpen
  \bibfield  {author} {\bibinfo {author} {\bibfnamefont {K.}~\bibnamefont
  {Ichiki}}, \bibinfo {author} {\bibfnamefont {K.}~\bibnamefont {Mimura}},
  \bibinfo {author} {\bibfnamefont {H.}~\bibnamefont {Anzai}}, \bibinfo
  {author} {\bibfnamefont {T.}~\bibnamefont {Uozumi}}, \bibinfo {author}
  {\bibfnamefont {H.}~\bibnamefont {Sato}}, \bibinfo {author} {\bibfnamefont
  {Y.}~\bibnamefont {Utsumi}}, \bibinfo {author} {\bibfnamefont
  {S.}~\bibnamefont {Ueda}}, \bibinfo {author} {\bibfnamefont {A.}~\bibnamefont
  {Mitsuda}}, \bibinfo {author} {\bibfnamefont {H.}~\bibnamefont {Wada}},
  \bibinfo {author} {\bibfnamefont {Y.}~\bibnamefont {Taguchi}}, \bibinfo
  {author} {\bibfnamefont {K.}~\bibnamefont {Shimada}}, \bibinfo {author}
  {\bibfnamefont {H.}~\bibnamefont {Namatame}}, \ and\ \bibinfo {author}
  {\bibfnamefont {M.}~\bibnamefont {Taniguchi}},\ }\href {\doibase
  10.1103/physrevb.96.045106} {\bibfield  {journal} {\bibinfo  {journal}
  {Physical Review B}\ }\textbf {\bibinfo {volume} {96}},\ \bibinfo {pages}
  {045106} (\bibinfo {year} {2017})}\BibitemShut {NoStop}%
\bibitem [{\citenamefont {Yokoyama}\ \emph {et~al.}(2019)\citenamefont
  {Yokoyama}, \citenamefont {Kawakami}, \citenamefont {Hirata}, \citenamefont
  {Takubo}, \citenamefont {Yamamoto}, \citenamefont {Abe}, \citenamefont
  {Mitsuda}, \citenamefont {Wada}, \citenamefont {Uozumi}, \citenamefont
  {Yamamoto}, \citenamefont {Matsuda}, \citenamefont {Kimura}, \citenamefont
  {Mimura},\ and\ \citenamefont {Wadati}}]{Yokoyama2019}%
  \BibitemOpen
  \bibfield  {author} {\bibinfo {author} {\bibfnamefont {Y.}~\bibnamefont
  {Yokoyama}}, \bibinfo {author} {\bibfnamefont {K.}~\bibnamefont {Kawakami}},
  \bibinfo {author} {\bibfnamefont {Y.}~\bibnamefont {Hirata}}, \bibinfo
  {author} {\bibfnamefont {K.}~\bibnamefont {Takubo}}, \bibinfo {author}
  {\bibfnamefont {K.}~\bibnamefont {Yamamoto}}, \bibinfo {author}
  {\bibfnamefont {K.}~\bibnamefont {Abe}}, \bibinfo {author} {\bibfnamefont
  {A.}~\bibnamefont {Mitsuda}}, \bibinfo {author} {\bibfnamefont
  {H.}~\bibnamefont {Wada}}, \bibinfo {author} {\bibfnamefont {T.}~\bibnamefont
  {Uozumi}}, \bibinfo {author} {\bibfnamefont {S.}~\bibnamefont {Yamamoto}},
  \bibinfo {author} {\bibfnamefont {I.}~\bibnamefont {Matsuda}}, \bibinfo
  {author} {\bibfnamefont {S.}~\bibnamefont {Kimura}}, \bibinfo {author}
  {\bibfnamefont {K.}~\bibnamefont {Mimura}}, \ and\ \bibinfo {author}
  {\bibfnamefont {H.}~\bibnamefont {Wadati}},\ }\href {\doibase
  10.1103/physrevb.100.115123} {\bibfield  {journal} {\bibinfo  {journal}
  {Physical Review B}\ }\textbf {\bibinfo {volume} {100}},\ \bibinfo {pages}
  {115123} (\bibinfo {year} {2019})}\BibitemShut {NoStop}%
\bibitem [{\citenamefont {Adams}\ \emph {et~al.}(1991)\citenamefont {Adams},
  \citenamefont {Heath}, \citenamefont {Jhans}, \citenamefont {Norman},\ and\
  \citenamefont {Leonard}}]{Adams1991}%
  \BibitemOpen
  \bibfield  {author} {\bibinfo {author} {\bibfnamefont {D.~M.}\ \bibnamefont
  {Adams}}, \bibinfo {author} {\bibfnamefont {A.~E.}\ \bibnamefont {Heath}},
  \bibinfo {author} {\bibfnamefont {H.}~\bibnamefont {Jhans}}, \bibinfo
  {author} {\bibfnamefont {A.}~\bibnamefont {Norman}}, \ and\ \bibinfo {author}
  {\bibfnamefont {S.}~\bibnamefont {Leonard}},\ }\href {\doibase
  10.1088/0953-8984/3/29/001} {\bibfield  {journal} {\bibinfo  {journal}
  {Journal of Physics: Condensed Matter}\ }\textbf {\bibinfo {volume} {3}},\
  \bibinfo {pages} {5465} (\bibinfo {year} {1991})}\BibitemShut {NoStop}%
\bibitem [{\citenamefont {Mitsuda}\ \emph {et~al.}(1997)\citenamefont
  {Mitsuda}, \citenamefont {Wada}, \citenamefont {Shiga}, \citenamefont
  {Katori},\ and\ \citenamefont {Goto}}]{Mitsuda1997}%
  \BibitemOpen
  \bibfield  {author} {\bibinfo {author} {\bibfnamefont {A.}~\bibnamefont
  {Mitsuda}}, \bibinfo {author} {\bibfnamefont {H.}~\bibnamefont {Wada}},
  \bibinfo {author} {\bibfnamefont {M.}~\bibnamefont {Shiga}}, \bibinfo
  {author} {\bibfnamefont {H.~A.}\ \bibnamefont {Katori}}, \ and\ \bibinfo
  {author} {\bibfnamefont {T.}~\bibnamefont {Goto}},\ }\href {\doibase
  10.1103/physrevb.55.12474} {\bibfield  {journal} {\bibinfo  {journal}
  {Physical Review B}\ }\textbf {\bibinfo {volume} {55}},\ \bibinfo {pages}
  {12474} (\bibinfo {year} {1997})}\BibitemShut {NoStop}%
\bibitem [{\citenamefont {Iyer}\ \emph {et~al.}(2018)\citenamefont {Iyer},
  \citenamefont {Basu}, \citenamefont {Paulose},\ and\ \citenamefont
  {Sampathkumaran}}]{Iyer2018}%
  \BibitemOpen
  \bibfield  {author} {\bibinfo {author} {\bibfnamefont {K.~K.}\ \bibnamefont
  {Iyer}}, \bibinfo {author} {\bibfnamefont {T.}~\bibnamefont {Basu}}, \bibinfo
  {author} {\bibfnamefont {P.}~\bibnamefont {Paulose}}, \ and\ \bibinfo
  {author} {\bibfnamefont {E.}~\bibnamefont {Sampathkumaran}},\ }\href
  {\doibase 10.1016/j.jmmm.2018.06.039} {\bibfield  {journal} {\bibinfo
  {journal} {Journal of Magnetism and Magnetic Materials}\ }\textbf {\bibinfo
  {volume} {465}},\ \bibinfo {pages} {515} (\bibinfo {year}
  {2018})}\BibitemShut {NoStop}%
\bibitem [{\citenamefont {K{\"o}lsch}\ \emph {et~al.}(2022)\citenamefont
  {K{\"o}lsch}, \citenamefont {Schuck}, \citenamefont {Huth}, \citenamefont
  {Fedchenko}, \citenamefont {Vasilyev}, \citenamefont {Chernov}, \citenamefont
  {Tkach}, \citenamefont {Elmers}, \citenamefont {Schö{\"o}nhense},
  \citenamefont {Schl{\"u}ter}, \citenamefont {Peixoto}, \citenamefont
  {Gloskowski},\ and\ \citenamefont {Krellner}}]{Koelsch2022}%
  \BibitemOpen
  \bibfield  {author} {\bibinfo {author} {\bibfnamefont {S.}~\bibnamefont
  {K{\"o}lsch}}, \bibinfo {author} {\bibfnamefont {A.}~\bibnamefont {Schuck}},
  \bibinfo {author} {\bibfnamefont {M.}~\bibnamefont {Huth}}, \bibinfo {author}
  {\bibfnamefont {O.}~\bibnamefont {Fedchenko}}, \bibinfo {author}
  {\bibfnamefont {D.}~\bibnamefont {Vasilyev}}, \bibinfo {author}
  {\bibfnamefont {S.}~\bibnamefont {Chernov}}, \bibinfo {author} {\bibfnamefont
  {O.}~\bibnamefont {Tkach}}, \bibinfo {author} {\bibfnamefont {H.-J.}\
  \bibnamefont {Elmers}}, \bibinfo {author} {\bibfnamefont {G.}~\bibnamefont
  {Schö{\"o}nhense}}, \bibinfo {author} {\bibfnamefont {C.}~\bibnamefont
  {Schl{\"u}ter}}, \bibinfo {author} {\bibfnamefont {T.~R.~F.}\ \bibnamefont
  {Peixoto}}, \bibinfo {author} {\bibfnamefont {A.}~\bibnamefont {Gloskowski}},
  \ and\ \bibinfo {author} {\bibfnamefont {C.}~\bibnamefont {Krellner}},\
  }\href {\doibase 10.1103/physrevmaterials.6.115003} {\bibfield  {journal}
  {\bibinfo  {journal} {Physical Review Materials}\ }\textbf {\bibinfo {volume}
  {6}},\ \bibinfo {pages} {115003} (\bibinfo {year} {2022})}\BibitemShut
  {NoStop}%
\bibitem [{\citenamefont {M{\aa}rtensson}\ \emph {et~al.}(1982)\citenamefont
  {M{\aa}rtensson}, \citenamefont {Reihl}, \citenamefont {Schneider},
  \citenamefont {Murgai}, \citenamefont {Gupta},\ and\ \citenamefont
  {Parks}}]{Maartensson1982}%
  \BibitemOpen
  \bibfield  {author} {\bibinfo {author} {\bibfnamefont {N.}~\bibnamefont
  {M{\aa}rtensson}}, \bibinfo {author} {\bibfnamefont {B.}~\bibnamefont
  {Reihl}}, \bibinfo {author} {\bibfnamefont {W.~D.}\ \bibnamefont
  {Schneider}}, \bibinfo {author} {\bibfnamefont {V.}~\bibnamefont {Murgai}},
  \bibinfo {author} {\bibfnamefont {L.~C.}\ \bibnamefont {Gupta}}, \ and\
  \bibinfo {author} {\bibfnamefont {R.~D.}\ \bibnamefont {Parks}},\ }\href
  {\doibase 10.1103/physrevb.25.1446} {\bibfield  {journal} {\bibinfo
  {journal} {Physical Review B}\ }\textbf {\bibinfo {volume} {25}},\ \bibinfo
  {pages} {1446} (\bibinfo {year} {1982})}\BibitemShut {NoStop}%
\bibitem [{\citenamefont {Mimura}\ \emph
  {et~al.}(2004{\natexlab{a}})\citenamefont {Mimura}, \citenamefont {Taguchi},
  \citenamefont {Fukuda}, \citenamefont {Mitsuda}, \citenamefont {Sakurai},
  \citenamefont {Ichikawa},\ and\ \citenamefont {Aita}}]{Mimura2004}%
  \BibitemOpen
  \bibfield  {author} {\bibinfo {author} {\bibfnamefont {K.}~\bibnamefont
  {Mimura}}, \bibinfo {author} {\bibfnamefont {Y.}~\bibnamefont {Taguchi}},
  \bibinfo {author} {\bibfnamefont {S.}~\bibnamefont {Fukuda}}, \bibinfo
  {author} {\bibfnamefont {A.}~\bibnamefont {Mitsuda}}, \bibinfo {author}
  {\bibfnamefont {J.}~\bibnamefont {Sakurai}}, \bibinfo {author} {\bibfnamefont
  {K.}~\bibnamefont {Ichikawa}}, \ and\ \bibinfo {author} {\bibfnamefont
  {O.}~\bibnamefont {Aita}},\ }\href {\doibase 10.1016/j.physb.2004.06.030}
  {\bibfield  {journal} {\bibinfo  {journal} {Physica B: Condensed Matter}\
  }\textbf {\bibinfo {volume} {351}},\ \bibinfo {pages} {292} (\bibinfo {year}
  {2004}{\natexlab{a}})}\BibitemShut {NoStop}%
\bibitem [{\citenamefont {Mimura}\ \emph
  {et~al.}(2004{\natexlab{b}})\citenamefont {Mimura}, \citenamefont {Taguchi},
  \citenamefont {Fukuda}, \citenamefont {Mitsuda}, \citenamefont {Sakurai},
  \citenamefont {Ichikawa},\ and\ \citenamefont {Aita}}]{Mimura2004a}%
  \BibitemOpen
  \bibfield  {author} {\bibinfo {author} {\bibfnamefont {K.}~\bibnamefont
  {Mimura}}, \bibinfo {author} {\bibfnamefont {Y.}~\bibnamefont {Taguchi}},
  \bibinfo {author} {\bibfnamefont {S.}~\bibnamefont {Fukuda}}, \bibinfo
  {author} {\bibfnamefont {A.}~\bibnamefont {Mitsuda}}, \bibinfo {author}
  {\bibfnamefont {J.}~\bibnamefont {Sakurai}}, \bibinfo {author} {\bibfnamefont
  {K.}~\bibnamefont {Ichikawa}}, \ and\ \bibinfo {author} {\bibfnamefont
  {O.}~\bibnamefont {Aita}},\ }\href {\doibase 10.1016/j.elspec.2004.02.098}
  {\bibfield  {journal} {\bibinfo  {journal} {Journal of Electron Spectroscopy
  and Related Phenomena}\ }\textbf {\bibinfo {volume} {137-140}},\ \bibinfo
  {pages} {529} (\bibinfo {year} {2004}{\natexlab{b}})}\BibitemShut {NoStop}%
\bibitem [{\citenamefont {Gray}\ \emph {et~al.}(2011)\citenamefont {Gray},
  \citenamefont {Papp}, \citenamefont {Ueda}, \citenamefont {Balke},
  \citenamefont {Yamashita}, \citenamefont {Plucinski}, \citenamefont
  {Min{\'{a}}r}, \citenamefont {Braun}, \citenamefont {Ylvisaker},
  \citenamefont {Schneider}, \citenamefont {Pickett}, \citenamefont {Ebert},
  \citenamefont {Kobayashi},\ and\ \citenamefont {Fadley}}]{Gray2011}%
  \BibitemOpen
  \bibfield  {author} {\bibinfo {author} {\bibfnamefont {A.~X.}\ \bibnamefont
  {Gray}}, \bibinfo {author} {\bibfnamefont {C.}~\bibnamefont {Papp}}, \bibinfo
  {author} {\bibfnamefont {S.}~\bibnamefont {Ueda}}, \bibinfo {author}
  {\bibfnamefont {B.}~\bibnamefont {Balke}}, \bibinfo {author} {\bibfnamefont
  {Y.}~\bibnamefont {Yamashita}}, \bibinfo {author} {\bibfnamefont
  {L.}~\bibnamefont {Plucinski}}, \bibinfo {author} {\bibfnamefont
  {J.}~\bibnamefont {Min{\'{a}}r}}, \bibinfo {author} {\bibfnamefont
  {J.}~\bibnamefont {Braun}}, \bibinfo {author} {\bibfnamefont {E.~R.}\
  \bibnamefont {Ylvisaker}}, \bibinfo {author} {\bibfnamefont {C.~M.}\
  \bibnamefont {Schneider}}, \bibinfo {author} {\bibfnamefont {W.~E.}\
  \bibnamefont {Pickett}}, \bibinfo {author} {\bibfnamefont {H.}~\bibnamefont
  {Ebert}}, \bibinfo {author} {\bibfnamefont {K.}~\bibnamefont {Kobayashi}}, \
  and\ \bibinfo {author} {\bibfnamefont {C.~S.}\ \bibnamefont {Fadley}},\
  }\href {\doibase 10.1038/nmat3089} {\bibfield  {journal} {\bibinfo  {journal}
  {Nature Materials}\ }\textbf {\bibinfo {volume} {10}},\ \bibinfo {pages}
  {759} (\bibinfo {year} {2011})}\BibitemShut {NoStop}%
\bibitem [{\citenamefont {Song}\ \emph {et~al.}(2023)\citenamefont {Song},
  \citenamefont {Schulz}, \citenamefont {Kliemt}, \citenamefont {Krellner},\
  and\ \citenamefont {Valent{\'{\i}}}}]{Song2023}%
  \BibitemOpen
  \bibfield  {author} {\bibinfo {author} {\bibfnamefont {Y.-J.}\ \bibnamefont
  {Song}}, \bibinfo {author} {\bibfnamefont {S.}~\bibnamefont {Schulz}},
  \bibinfo {author} {\bibfnamefont {K.}~\bibnamefont {Kliemt}}, \bibinfo
  {author} {\bibfnamefont {C.}~\bibnamefont {Krellner}}, \ and\ \bibinfo
  {author} {\bibfnamefont {R.}~\bibnamefont {Valent{\'{\i}}}},\ }\href
  {\doibase 10.1103/physrevb.107.075149} {\bibfield  {journal} {\bibinfo
  {journal} {Physical Review B}\ }\textbf {\bibinfo {volume} {107}},\ \bibinfo
  {pages} {075149} (\bibinfo {year} {2023})}\BibitemShut {NoStop}%
\bibitem [{\citenamefont {Kliemt}\ \emph {et~al.}(2022)\citenamefont {Kliemt},
  \citenamefont {Peters}, \citenamefont {Reiser}, \citenamefont {Ocker},
  \citenamefont {Walther}, \citenamefont {Tran}, \citenamefont {Cho},
  \citenamefont {Merz}, \citenamefont {Haghighirad}, \citenamefont {Hezel},
  \citenamefont {Ritter},\ and\ \citenamefont {Krellner}}]{Kliemt2022}%
  \BibitemOpen
  \bibfield  {author} {\bibinfo {author} {\bibfnamefont {K.}~\bibnamefont
  {Kliemt}}, \bibinfo {author} {\bibfnamefont {M.}~\bibnamefont {Peters}},
  \bibinfo {author} {\bibfnamefont {I.}~\bibnamefont {Reiser}}, \bibinfo
  {author} {\bibfnamefont {M.}~\bibnamefont {Ocker}}, \bibinfo {author}
  {\bibfnamefont {F.}~\bibnamefont {Walther}}, \bibinfo {author} {\bibfnamefont
  {D.-M.}\ \bibnamefont {Tran}}, \bibinfo {author} {\bibfnamefont
  {E.}~\bibnamefont {Cho}}, \bibinfo {author} {\bibfnamefont {M.}~\bibnamefont
  {Merz}}, \bibinfo {author} {\bibfnamefont {A.~A.}\ \bibnamefont
  {Haghighirad}}, \bibinfo {author} {\bibfnamefont {D.~C.}\ \bibnamefont
  {Hezel}}, \bibinfo {author} {\bibfnamefont {F.}~\bibnamefont {Ritter}}, \
  and\ \bibinfo {author} {\bibfnamefont {C.}~\bibnamefont {Krellner}},\ }\href
  {\doibase 10.1021/acs.cgd.2c00475} {\bibfield  {journal} {\bibinfo  {journal}
  {Crystal Growth and Design}\ }\textbf {\bibinfo {volume} {22}},\ \bibinfo
  {pages} {5399} (\bibinfo {year} {2022})}\BibitemShut {NoStop}%
\bibitem [{\citenamefont {Schlueter}\ \emph {et~al.}(2019)\citenamefont
  {Schlueter}, \citenamefont {Gloskovskii}, \citenamefont {Ederer},
  \citenamefont {Schostak}, \citenamefont {Piec}, \citenamefont {Sarkar},
  \citenamefont {Matveyev}, \citenamefont {Lömker}, \citenamefont {Sing},
  \citenamefont {Claessen}, \citenamefont {Wiemann}, \citenamefont {Schneider},
  \citenamefont {Medjanik}, \citenamefont {Schönhense}, \citenamefont {Amann},
  \citenamefont {Nilsson},\ and\ \citenamefont {Drube}}]{Schlueter2019}%
  \BibitemOpen
  \bibfield  {author} {\bibinfo {author} {\bibfnamefont {C.}~\bibnamefont
  {Schlueter}}, \bibinfo {author} {\bibfnamefont {A.}~\bibnamefont
  {Gloskovskii}}, \bibinfo {author} {\bibfnamefont {K.}~\bibnamefont {Ederer}},
  \bibinfo {author} {\bibfnamefont {I.}~\bibnamefont {Schostak}}, \bibinfo
  {author} {\bibfnamefont {S.}~\bibnamefont {Piec}}, \bibinfo {author}
  {\bibfnamefont {I.}~\bibnamefont {Sarkar}}, \bibinfo {author} {\bibfnamefont
  {Y.}~\bibnamefont {Matveyev}}, \bibinfo {author} {\bibfnamefont
  {P.}~\bibnamefont {Lömker}}, \bibinfo {author} {\bibfnamefont
  {M.}~\bibnamefont {Sing}}, \bibinfo {author} {\bibfnamefont {R.}~\bibnamefont
  {Claessen}}, \bibinfo {author} {\bibfnamefont {C.}~\bibnamefont {Wiemann}},
  \bibinfo {author} {\bibfnamefont {C.~M.}\ \bibnamefont {Schneider}}, \bibinfo
  {author} {\bibfnamefont {K.}~\bibnamefont {Medjanik}}, \bibinfo {author}
  {\bibfnamefont {G.}~\bibnamefont {Schönhense}}, \bibinfo {author}
  {\bibfnamefont {P.}~\bibnamefont {Amann}}, \bibinfo {author} {\bibfnamefont
  {A.}~\bibnamefont {Nilsson}}, \ and\ \bibinfo {author} {\bibfnamefont
  {W.}~\bibnamefont {Drube}},\ }in\ \href {\doibase 10.1063/1.5084611} {\emph
  {\bibinfo {booktitle} {{AIP} Conference Proceedings}}}\ (\bibinfo
  {publisher} {Author(s)},\ \bibinfo {year} {2019})\BibitemShut {NoStop}%
\bibitem [{\citenamefont {Medjanik}\ \emph {et~al.}(2017)\citenamefont
  {Medjanik}, \citenamefont {Fedchenko}, \citenamefont {Chernov}, \citenamefont
  {Kutnyakhov}, \citenamefont {Ellguth}, \citenamefont {Oelsner}, \citenamefont
  {Sch{\"o}nhense}, \citenamefont {Peixoto}, \citenamefont {Lutz},
  \citenamefont {Min}, \citenamefont {Reinert}, \citenamefont {Däster},
  \citenamefont {Acremann}, \citenamefont {Viefhaus}, \citenamefont {Wurth},
  \citenamefont {Elmers},\ and\ \citenamefont {Sch{\"o}nhense}}]{Medjanik2017}%
  \BibitemOpen
  \bibfield  {author} {\bibinfo {author} {\bibfnamefont {K.}~\bibnamefont
  {Medjanik}}, \bibinfo {author} {\bibfnamefont {O.}~\bibnamefont {Fedchenko}},
  \bibinfo {author} {\bibfnamefont {S.}~\bibnamefont {Chernov}}, \bibinfo
  {author} {\bibfnamefont {D.}~\bibnamefont {Kutnyakhov}}, \bibinfo {author}
  {\bibfnamefont {M.}~\bibnamefont {Ellguth}}, \bibinfo {author} {\bibfnamefont
  {A.}~\bibnamefont {Oelsner}}, \bibinfo {author} {\bibfnamefont
  {B.}~\bibnamefont {Sch{\"o}nhense}}, \bibinfo {author} {\bibfnamefont
  {T.~R.~F.}\ \bibnamefont {Peixoto}}, \bibinfo {author} {\bibfnamefont
  {P.}~\bibnamefont {Lutz}}, \bibinfo {author} {\bibfnamefont {C.-H.}\
  \bibnamefont {Min}}, \bibinfo {author} {\bibfnamefont {F.}~\bibnamefont
  {Reinert}}, \bibinfo {author} {\bibfnamefont {S.}~\bibnamefont {Däster}},
  \bibinfo {author} {\bibfnamefont {Y.}~\bibnamefont {Acremann}}, \bibinfo
  {author} {\bibfnamefont {J.}~\bibnamefont {Viefhaus}}, \bibinfo {author}
  {\bibfnamefont {W.}~\bibnamefont {Wurth}}, \bibinfo {author} {\bibfnamefont
  {H.~J.}\ \bibnamefont {Elmers}}, \ and\ \bibinfo {author} {\bibfnamefont
  {G.}~\bibnamefont {Sch{\"o}nhense}},\ }\href {\doibase 10.1038/nmat4875}
  {\bibfield  {journal} {\bibinfo  {journal} {Nature Materials}\ }\textbf
  {\bibinfo {volume} {16}},\ \bibinfo {pages} {615} (\bibinfo {year}
  {2017})}\BibitemShut {NoStop}%
\bibitem [{\citenamefont {Agustsson}\ \emph {et~al.}(2021)\citenamefont
  {Agustsson}, \citenamefont {Chernov}, \citenamefont {Medjanik}, \citenamefont
  {Babenkov}, \citenamefont {Fedchenko}, \citenamefont {Vasilyev},
  \citenamefont {Schlueter}, \citenamefont {Gloskovskii}, \citenamefont
  {Matveyev}, \citenamefont {Kliemt}, \citenamefont {Krellner}, \citenamefont
  {Demsar}, \citenamefont {Sch{\"o}nhense},\ and\ \citenamefont
  {Elmers}}]{Agustsson2021}%
  \BibitemOpen
  \bibfield  {author} {\bibinfo {author} {\bibfnamefont {S.~Y.}\ \bibnamefont
  {Agustsson}}, \bibinfo {author} {\bibfnamefont {S.~V.}\ \bibnamefont
  {Chernov}}, \bibinfo {author} {\bibfnamefont {K.}~\bibnamefont {Medjanik}},
  \bibinfo {author} {\bibfnamefont {S.}~\bibnamefont {Babenkov}}, \bibinfo
  {author} {\bibfnamefont {O.}~\bibnamefont {Fedchenko}}, \bibinfo {author}
  {\bibfnamefont {D.}~\bibnamefont {Vasilyev}}, \bibinfo {author}
  {\bibfnamefont {C.}~\bibnamefont {Schlueter}}, \bibinfo {author}
  {\bibfnamefont {A.}~\bibnamefont {Gloskovskii}}, \bibinfo {author}
  {\bibfnamefont {Y.}~\bibnamefont {Matveyev}}, \bibinfo {author}
  {\bibfnamefont {K.}~\bibnamefont {Kliemt}}, \bibinfo {author} {\bibfnamefont
  {C.}~\bibnamefont {Krellner}}, \bibinfo {author} {\bibfnamefont
  {J.}~\bibnamefont {Demsar}}, \bibinfo {author} {\bibfnamefont
  {G.}~\bibnamefont {Sch{\"o}nhense}}, \ and\ \bibinfo {author} {\bibfnamefont
  {H.-J.}\ \bibnamefont {Elmers}},\ }\href {\doibase 10.1088/1361-648x/abe479}
  {\bibfield  {journal} {\bibinfo  {journal} {Journal of Physics: Condensed
  Matter}\ }\textbf {\bibinfo {volume} {33}},\ \bibinfo {pages} {205601}
  (\bibinfo {year} {2021})}\BibitemShut {NoStop}%
\bibitem [{\citenamefont {Babenkov}\ \emph {et~al.}(2019)\citenamefont
  {Babenkov}, \citenamefont {Medjanik}, \citenamefont {Vasilyev}, \citenamefont
  {Chernov}, \citenamefont {Schlueter}, \citenamefont {Gloskovskii},
  \citenamefont {Matveyev}, \citenamefont {Drube}, \citenamefont {Schönhense},
  \citenamefont {Rossnagel}, \citenamefont {Elmers},\ and\ \citenamefont
  {Sch{\"o}nhense}}]{Babenkov2019}%
  \BibitemOpen
  \bibfield  {author} {\bibinfo {author} {\bibfnamefont {S.}~\bibnamefont
  {Babenkov}}, \bibinfo {author} {\bibfnamefont {K.}~\bibnamefont {Medjanik}},
  \bibinfo {author} {\bibfnamefont {D.}~\bibnamefont {Vasilyev}}, \bibinfo
  {author} {\bibfnamefont {S.}~\bibnamefont {Chernov}}, \bibinfo {author}
  {\bibfnamefont {C.}~\bibnamefont {Schlueter}}, \bibinfo {author}
  {\bibfnamefont {A.}~\bibnamefont {Gloskovskii}}, \bibinfo {author}
  {\bibfnamefont {Y.}~\bibnamefont {Matveyev}}, \bibinfo {author}
  {\bibfnamefont {W.}~\bibnamefont {Drube}}, \bibinfo {author} {\bibfnamefont
  {B.}~\bibnamefont {Schönhense}}, \bibinfo {author} {\bibfnamefont
  {K.}~\bibnamefont {Rossnagel}}, \bibinfo {author} {\bibfnamefont {H.-J.}\
  \bibnamefont {Elmers}}, \ and\ \bibinfo {author} {\bibfnamefont
  {G.}~\bibnamefont {Sch{\"o}nhense}},\ }\href {\doibase
  10.1038/s42005-019-0208-7} {\bibfield  {journal} {\bibinfo  {journal}
  {Communications Physics}\ }\textbf {\bibinfo {volume} {2}} (\bibinfo {year}
  {2019}),\ 10.1038/s42005-019-0208-7}\BibitemShut {NoStop}%
\bibitem [{\citenamefont {Medjanik}\ \emph {et~al.}(2019)\citenamefont
  {Medjanik}, \citenamefont {Babenkov}, \citenamefont {Chernov}, \citenamefont
  {Vasilyev}, \citenamefont {Sch{\"o}nhense}, \citenamefont {Schlueter},
  \citenamefont {Gloskovskii}, \citenamefont {Matveyev}, \citenamefont {Drube},
  \citenamefont {Elmers},\ and\ \citenamefont {Sch{\"o}nhense}}]{Medjanik2019}%
  \BibitemOpen
  \bibfield  {author} {\bibinfo {author} {\bibfnamefont {K.}~\bibnamefont
  {Medjanik}}, \bibinfo {author} {\bibfnamefont {S.~V.}\ \bibnamefont
  {Babenkov}}, \bibinfo {author} {\bibfnamefont {S.}~\bibnamefont {Chernov}},
  \bibinfo {author} {\bibfnamefont {D.}~\bibnamefont {Vasilyev}}, \bibinfo
  {author} {\bibfnamefont {B.}~\bibnamefont {Sch{\"o}nhense}}, \bibinfo
  {author} {\bibfnamefont {C.}~\bibnamefont {Schlueter}}, \bibinfo {author}
  {\bibfnamefont {A.}~\bibnamefont {Gloskovskii}}, \bibinfo {author}
  {\bibfnamefont {Y.}~\bibnamefont {Matveyev}}, \bibinfo {author}
  {\bibfnamefont {W.}~\bibnamefont {Drube}}, \bibinfo {author} {\bibfnamefont
  {H.~J.}\ \bibnamefont {Elmers}}, \ and\ \bibinfo {author} {\bibfnamefont
  {G.}~\bibnamefont {Sch{\"o}nhense}},\ }\href {\doibase
  10.1107/s1600577519012773} {\bibfield  {journal} {\bibinfo  {journal}
  {Journal of Synchrotron Radiation}\ }\textbf {\bibinfo {volume} {26}},\
  \bibinfo {pages} {1996} (\bibinfo {year} {2019})}\BibitemShut {NoStop}%
\bibitem [{\citenamefont {Seah}\ and\ \citenamefont {Dench}(1979)}]{Seah1979}%
  \BibitemOpen
  \bibfield  {author} {\bibinfo {author} {\bibfnamefont {M.~P.}\ \bibnamefont
  {Seah}}\ and\ \bibinfo {author} {\bibfnamefont {W.~A.}\ \bibnamefont
  {Dench}},\ }\href {\doibase 10.1002/sia.740010103} {\bibfield  {journal}
  {\bibinfo  {journal} {Surface and Interface Analysis}\ }\textbf {\bibinfo
  {volume} {1}},\ \bibinfo {pages} {2} (\bibinfo {year} {1979})}\BibitemShut
  {NoStop}%
\bibitem [{\citenamefont {Mimura}\ \emph {et~al.}(2011)\citenamefont {Mimura},
  \citenamefont {Uozumi}, \citenamefont {Ishizu}, \citenamefont {Motonami},
  \citenamefont {Sato}, \citenamefont {Utsumi}, \citenamefont {Ueda},
  \citenamefont {Mitsuda}, \citenamefont {Shimada}, \citenamefont {Taguchi},
  \citenamefont {Yamashita}, \citenamefont {Yoshikawa}, \citenamefont
  {Namatame}, \citenamefont {Taniguchi},\ and\ \citenamefont
  {Kobayashi}}]{Mimura2011}%
  \BibitemOpen
  \bibfield  {author} {\bibinfo {author} {\bibfnamefont {K.}~\bibnamefont
  {Mimura}}, \bibinfo {author} {\bibfnamefont {T.}~\bibnamefont {Uozumi}},
  \bibinfo {author} {\bibfnamefont {T.}~\bibnamefont {Ishizu}}, \bibinfo
  {author} {\bibfnamefont {S.}~\bibnamefont {Motonami}}, \bibinfo {author}
  {\bibfnamefont {H.}~\bibnamefont {Sato}}, \bibinfo {author} {\bibfnamefont
  {Y.}~\bibnamefont {Utsumi}}, \bibinfo {author} {\bibfnamefont
  {S.}~\bibnamefont {Ueda}}, \bibinfo {author} {\bibfnamefont {A.}~\bibnamefont
  {Mitsuda}}, \bibinfo {author} {\bibfnamefont {K.}~\bibnamefont {Shimada}},
  \bibinfo {author} {\bibfnamefont {Y.}~\bibnamefont {Taguchi}}, \bibinfo
  {author} {\bibfnamefont {Y.}~\bibnamefont {Yamashita}}, \bibinfo {author}
  {\bibfnamefont {H.}~\bibnamefont {Yoshikawa}}, \bibinfo {author}
  {\bibfnamefont {H.}~\bibnamefont {Namatame}}, \bibinfo {author}
  {\bibfnamefont {M.}~\bibnamefont {Taniguchi}}, \ and\ \bibinfo {author}
  {\bibfnamefont {K.}~\bibnamefont {Kobayashi}},\ }\href {\doibase
  10.1143/jjap.50.05fd03} {\bibfield  {journal} {\bibinfo  {journal} {Japanese
  Journal of Applied Physics}\ }\textbf {\bibinfo {volume} {50}},\ \bibinfo
  {pages} {05FD03} (\bibinfo {year} {2011})}\BibitemShut {NoStop}%
\bibitem [{\citenamefont {Haverkort}(2016)}]{Haverkort2016}%
  \BibitemOpen
  \bibfield  {author} {\bibinfo {author} {\bibfnamefont {M.~W.}\ \bibnamefont
  {Haverkort}},\ }\href {\doibase 10.1088/1742-6596/712/1/012001} {\bibfield
  {journal} {\bibinfo  {journal} {Journal of Physics: Conference Series}\
  }\textbf {\bibinfo {volume} {712}},\ \bibinfo {pages} {012001} (\bibinfo
  {year} {2016})}\BibitemShut {NoStop}%
\bibitem [{\citenamefont {Usachov}\ \emph {et~al.}(2020)\citenamefont
  {Usachov}, \citenamefont {Tarasov}, \citenamefont {Schulz}, \citenamefont
  {Bokai}, \citenamefont {Tupitsyn}, \citenamefont {Poelchen}, \citenamefont
  {Seiro}, \citenamefont {Caroca-Canales}, \citenamefont {Kliemt},
  \citenamefont {Mende}, \citenamefont {Kummer}, \citenamefont {Krellner},
  \citenamefont {Muntwiler}, \citenamefont {Li}, \citenamefont {Laubschat},
  \citenamefont {Geibel}, \citenamefont {Chulkov}, \citenamefont {Fujimori},\
  and\ \citenamefont {Vyalikh}}]{Usachov2020}%
  \BibitemOpen
  \bibfield  {author} {\bibinfo {author} {\bibfnamefont {D.~Y.}\ \bibnamefont
  {Usachov}}, \bibinfo {author} {\bibfnamefont {A.~V.}\ \bibnamefont
  {Tarasov}}, \bibinfo {author} {\bibfnamefont {S.}~\bibnamefont {Schulz}},
  \bibinfo {author} {\bibfnamefont {K.~A.}\ \bibnamefont {Bokai}}, \bibinfo
  {author} {\bibfnamefont {I.~I.}\ \bibnamefont {Tupitsyn}}, \bibinfo {author}
  {\bibfnamefont {G.}~\bibnamefont {Poelchen}}, \bibinfo {author}
  {\bibfnamefont {S.}~\bibnamefont {Seiro}}, \bibinfo {author} {\bibfnamefont
  {N.}~\bibnamefont {Caroca-Canales}}, \bibinfo {author} {\bibfnamefont
  {K.}~\bibnamefont {Kliemt}}, \bibinfo {author} {\bibfnamefont
  {M.}~\bibnamefont {Mende}}, \bibinfo {author} {\bibfnamefont
  {K.}~\bibnamefont {Kummer}}, \bibinfo {author} {\bibfnamefont
  {C.}~\bibnamefont {Krellner}}, \bibinfo {author} {\bibfnamefont
  {M.}~\bibnamefont {Muntwiler}}, \bibinfo {author} {\bibfnamefont
  {H.}~\bibnamefont {Li}}, \bibinfo {author} {\bibfnamefont {C.}~\bibnamefont
  {Laubschat}}, \bibinfo {author} {\bibfnamefont {C.}~\bibnamefont {Geibel}},
  \bibinfo {author} {\bibfnamefont {E.~V.}\ \bibnamefont {Chulkov}}, \bibinfo
  {author} {\bibfnamefont {S.~I.}\ \bibnamefont {Fujimori}}, \ and\ \bibinfo
  {author} {\bibfnamefont {D.~V.}\ \bibnamefont {Vyalikh}},\ }\href {\doibase
  10.1103/physrevb.102.205102} {\bibfield  {journal} {\bibinfo  {journal}
  {Physical Review B}\ }\textbf {\bibinfo {volume} {102}},\ \bibinfo {pages}
  {205102} (\bibinfo {year} {2020})}\BibitemShut {NoStop}%
\bibitem [{\citenamefont {Koepernik}\ and\ \citenamefont
  {Eschrig}(1999)}]{Koepernik1999}%
  \BibitemOpen
  \bibfield  {author} {\bibinfo {author} {\bibfnamefont {K.}~\bibnamefont
  {Koepernik}}\ and\ \bibinfo {author} {\bibfnamefont {H.}~\bibnamefont
  {Eschrig}},\ }\href {\doibase 10.1103/physrevb.59.1743} {\bibfield  {journal}
  {\bibinfo  {journal} {Physical Review B}\ }\textbf {\bibinfo {volume} {59}},\
  \bibinfo {pages} {1743} (\bibinfo {year} {1999})}\BibitemShut {NoStop}%
\bibitem [{\citenamefont {Opahle}\ \emph {et~al.}(1999)\citenamefont {Opahle},
  \citenamefont {Koepernik},\ and\ \citenamefont {Eschrig}}]{Opahle1999}%
  \BibitemOpen
  \bibfield  {author} {\bibinfo {author} {\bibfnamefont {I.}~\bibnamefont
  {Opahle}}, \bibinfo {author} {\bibfnamefont {K.}~\bibnamefont {Koepernik}}, \
  and\ \bibinfo {author} {\bibfnamefont {H.}~\bibnamefont {Eschrig}},\ }\href
  {\doibase 10.1103/physrevb.60.14035} {\bibfield  {journal} {\bibinfo
  {journal} {Physical Review B}\ }\textbf {\bibinfo {volume} {60}},\ \bibinfo
  {pages} {14035} (\bibinfo {year} {1999})}\BibitemShut {NoStop}%
\bibitem [{\citenamefont {Mitsuda}\ \emph {et~al.}(2000)\citenamefont
  {Mitsuda}, \citenamefont {Wada}, \citenamefont {Shiga},\ and\ \citenamefont
  {Tanaka}}]{Mitsuda2000}%
  \BibitemOpen
  \bibfield  {author} {\bibinfo {author} {\bibfnamefont {A.}~\bibnamefont
  {Mitsuda}}, \bibinfo {author} {\bibfnamefont {H.}~\bibnamefont {Wada}},
  \bibinfo {author} {\bibfnamefont {M.}~\bibnamefont {Shiga}}, \ and\ \bibinfo
  {author} {\bibfnamefont {T.}~\bibnamefont {Tanaka}},\ }\href {\doibase
  10.1088/0953-8984/12/24/317} {\bibfield  {journal} {\bibinfo  {journal}
  {Journal of Physics: Condensed Matter}\ }\textbf {\bibinfo {volume} {12}},\
  \bibinfo {pages} {5287} (\bibinfo {year} {2000})}\BibitemShut {NoStop}%
\bibitem [{\citenamefont {Palenzona}\ \emph {et~al.}(1987)\citenamefont
  {Palenzona}, \citenamefont {Cirafici},\ and\ \citenamefont
  {Canepa}}]{Palenzona1987}%
  \BibitemOpen
  \bibfield  {author} {\bibinfo {author} {\bibfnamefont {A.}~\bibnamefont
  {Palenzona}}, \bibinfo {author} {\bibfnamefont {S.}~\bibnamefont {Cirafici}},
  \ and\ \bibinfo {author} {\bibfnamefont {F.}~\bibnamefont {Canepa}},\ }\href
  {\doibase 10.1016/0022-5088(87)90479-6} {\bibfield  {journal} {\bibinfo
  {journal} {Journal of the Less Common Metals}\ }\textbf {\bibinfo {volume}
  {135}},\ \bibinfo {pages} {185} (\bibinfo {year} {1987})}\BibitemShut
  {NoStop}%
\end{thebibliography}
%

\end{document}